\documentclass[12pt,letterpaper]{article}  

\usepackage[includeheadfoot,
            marginratio={1:1,2:3}, 
            width=420pt, 
            height=688pt,]{geometry}

\usepackage{physics}
\usepackage{amsmath}
\usepackage{amsfonts}
\usepackage{amssymb}
\usepackage{mathtools}
\usepackage{graphicx}
\usepackage{tikz}
\usepackage{tikz,tcolorbox}
\usepackage{booktabs}
\usepackage{adjustbox}
\usepackage[utf8]{inputenc}
\usepackage[splitrule]{footmisc} 
\usepackage{slashed}
\usepackage{float}

\usepackage{empheq}
\usepackage{paralist}
\usepackage{cite}
\usepackage[hidelinks]{hyperref}
\usepackage{xcolor}
\usepackage{tensor}

\newcommand{\nc}{\newcommand}
\nc{\lb}{\llbracket}
\nc{\rb}{\rrbracket}
\nc{\gl}{\llbracket}
\nc{\gr}{\rrbracket}
\nc{\del}{\partial}

\nc{\eq}[1]{\begin{equation}
                     \begin{split} #1 \end{split}
                     \end{equation}}
\nc{\ul}{\underline}
\nc{\ov}{\overline}

\nc{\fa}{\hat}
\nc{\fb}{\MakeUppercase}
\nc{\fc}{\tilde}

\allowdisplaybreaks[2]
\numberwithin{equation}{section}

\begin{document}

\vspace*{-1.5cm}
\begin{flushright}
  {\small
  MPP-2023-96\\
  }
\end{flushright}

\vspace{1.0cm}
\begin{center}
  {\Large
    The Emergence Proposal and the Emergent String \\[0.3cm]
  } 
\vspace{0.4cm}

\end{center}

\vspace{0.25cm}
\begin{center}
Ralph Blumenhagen$^{1,2}$, Aleksandar Gligovic$^{1,2}$ and Antonia Paraskevopoulou$^{1,3}$
\end{center}

\vspace{0.0cm}
\begin{center} 
\emph{
$^{1}$ 
Max-Planck-Institut f\"ur Physik (Werner-Heisenberg-Institut), \\ 
F\"ohringer Ring 6,  80805 M\"unchen, Germany } 
\\[0.1cm] 
\vspace{0.25cm} 
\emph{$^{2}$ Exzellenzcluster ORIGINS, Boltzmannstr. 2, D-85748 Garching, Germany}\\[0.1cm]
\vspace{0.25cm} 
\emph{$^{3}$ Ludwig-Maximilians-Universit{\"a}t M\"unchen, Fakult{\"a}t f{\"u}r Physik,\\ 
Theresienstr.~37, 80333 M\"unchen, Germany}\\[0.1cm]
\vspace{0.3cm}
\end{center} 

\vspace{0.5cm}


\begin{abstract}
We explore   the Emergence Proposal for the moduli metric and the
gauge couplings in a concrete model with 7 saxionic and 7 axionic
moduli fields,
namely the compactification of the
type IIA superstring on a 6-dimensional toroidal orbifold.
We show that consistency requires integrating out precisely the 12
towers of light particle species
arising from KK and string/brane winding modes and
one asymptotically tensionless string up to the species scale.
After pointing out an issue with the correct definition of the species
scale in the presence of string towers, 
we carry  out the emergence computation and find
that the KK and winding modes indeed impose  the classical moduli dependence on the
one-loop corrections, while the emergent string induces 
moduli dependent logarithmic suppressions.
The interpretation of these results for  the Emergence Proposal are discussed
revealing a couple of new and still not completely settled aspects.
\end{abstract}

\clearpage

\tableofcontents

\section{Introduction}
\label{sec:intro}

Even though technically string theory is a well understood and applied
mathema-tical framework of quantum gravity, the recently
developed swampland program has revealed that, despite all
its technical beauty, it conceals some of its underlying physical
principles.
It is the aim of the swampland program (see
\cite{Palti:2019pca,vanBeest:2021lhn,Grana:2021zvf,Agmon:2022thq} for reviews)
to bring them to the front
and, in particular, to uncover  its consequences for our low-energy
effective field theory framework, which we are used to employ very
successfully, for instance, in the description of particle physics in the
form of the Standard Model.

One of the first  formulated  swampland conjectures is the Swampland Distance Conjecture \cite{Ooguri:2006in} (see
\cite{Baume:2016psm,Klaewer:2016kiy} for a refined version), which was studied in the context of here relevant $N=2$ supersymmetry in 4D in \cite{Grimm:2018ohb,Blumenhagen:2018nts,Grimm:2018cpv,Corvilain:2018lgw,Joshi:2019nzi,Marchesano:2019ifh,Font:2019cxq,Erkinger:2019umg,Lee:2019wij,Baume:2019sry}.
It has been generalized  to more specialized set-ups, like the AdS
Distance  Conjecture \cite{Lust:2019zwm} and the Gravitino Distance Conjecture \cite{Cribiori:2021gbf,Castellano:2021yye}.
The Swampland Distance Conjecture says that in asymptotic limits in field space towers of exponentially light particle species appear.
These are threatening the reliability of a  Wilsonian low-energy effective
field theory (EFT)  that is keeping only the  degrees of freedom below
a certain cut-off.
In some sense this is an awkward situation, as the stringy corrections
to such a low-energy description are usually believed to be only under control
in such asymptotic, i.e. weak coupling limits. And now another
effect of quantum gravity is about to threaten its validity in precisely
such regions or at least to restrict strongly its regime of validity?
By moving over trans-Planckian distances in field space, new modes
need to be integrated into the EFT which in principle could
change  certain couplings in the old EFT by inducing large corrections.

The so-called Emergence Proposal
\cite{Heidenreich:2017sim,Grimm:2018ohb,Heidenreich:2018kpg,
  Corvilain:2018lgw}
provides a new aspect to this situation.
It is based on a field theoretical
analysis  of the one-loop effects  the  light towers of states have  on certain
quantities in the low-energy effective action.
Initially, mostly  the kinetic terms for scalars, fermions and gauge
couplings  in four dimensions had been
considered, but more recently a more thorough analysis of emergence was
carried out in \cite{Castellano:2021mmx,Castellano:2022bvr}, which also considered arbitrary dimensions and also
e.g. scalar potentials. In \cite{Blumenhagen:2019qcg,Blumenhagen:2019vgj}
it was shown that the concept of
emergence can also be applied to the effective field theory
living in a strongly warped throat. This  appears close
to a conifold point in complex structure space and is at finite distance.

The important aspect of such a one-loop computation is that one
integrates out those states whose masses are below the cut-off
of quantum gravity, which naively is the four-dimensional Planck scale
that, however,  in the presence of many light species is known
to be lowered to the species scale \cite{Dvali:2007hz,Dvali:2007wp}.
Interestingly, there are two classes of arguments regarding the definition of the latter. In what we will call the QFT picture, it can be defined as the energy scale
where the one-loop corrections to graviton scattering processes
are of the same order as the tree-level ones. As such, it is defined by a one-loop effect leading to the intriguing
result that the so-defined sum over loop amplitudes turns
out to be independent of $\hbar$ and looks  like a classical
contribution. Alternatively, the species scale may be defined as the
inverse radius of the smallest Black Hole that the EFT can
describe. We will refer to this as the BH picture. The compatibility of those two definitions
is a tricky  issue and will be of relevance in the course of this work. Moreover, the species scale is moduli dependent which
was analyzed  more recently 
in \cite{vandeHeisteeg:2022btw,Cribiori:2022nke,vandeHeisteeg:2023ubh,Andriot:2023isc,vandeHeisteeg:2023uxj}.

Furthermore, it turned out that in simple examples
the moduli dependent induced field
metrics were of the same functional form as the usual tree-level ones.
This gave rise to the idea of two possible meanings of emergence
in quantum gravity. Following the formulation of \cite{Castellano:2022bvr}, the Strong
Emergence Proposal  \cite{Palti:2019pca} states
\begin{quotation}
  \noindent
  {\it  Strong Emergence: In a theory of Quantum Gravity all light particles in a perturbative
regime have no kinetic terms in the UV. The required kinetic terms appear as an IR effect
due to loop corrections involving the sum over a tower of massless states.}
\end{quotation}
This is a very far-reaching and general proposal which would mean that
quantum gravity in the UV is maybe a very simple theory, e.g. of purely topological 
nature, which only gives rise to geometric low-energy effective
theories by integrating out the light towers of states up to the
species scale.
See \cite{Castellano:2023qhp} for an application of this principle
to the Yukawa coupling in the Standard Model.
Since the Strong Emergence Proposal  is so general, it can in principle easily be falsified by
finding a controllable model that violates its claim.

A much milder version is the Weak Emergence Proposal
  which does not assume vanishing kinetic terms in the UV and makes a
  statement about quantum corrections to the metrics that match the
  ``tree level" behavior. To describe the objective and the results
  of this paper, it turns out to be useful to distinguish
   between two variants of Weak Emergence, called Variant A and Variant B. The more restrictive Variant A may be formulated as
\begin{quotation}
\noindent
  {\it Weak Emergence (Variant A): In a consistent theory of Quantum
    Gravity, for any singularity at infinite distance in the moduli
    space of the EFT, there are associated infinite towers of states
    becoming massless. These towers induce quantum corrections
    to the singular kinetic terms matching their
    tree-level behavior.}
\end{quotation}
\noindent
Implicitly this also means that no new singular dependence is
generated at one-loop.
This formulation restricts to the dependence of the
  kinetic terms on the single modulus taken to infinity.
  This is what was mostly studied so far in the literature.
  However, the full moduli metric will also depend on the many
  directions orthogonal to the asymptotic one and we would like
  to check whether also this dependence is (at least partially)
  recovered. 
   We therefore add a second variant of Weak Emergence, which reads
\begin{quotation}
  \noindent
  {\it 
  Weak Emergence (Variant B):  At infinite distance in
    the EFT moduli space, infinite towers of light states appear which
    induce quantum corrections matching, beyond the singular ones from
    Variant A, some or even all the orthogonal  kinetic terms. }
\end{quotation}
\noindent
Since the second variant talks also about the non-singular terms,
it is clearly stronger than the first
  and closer in spirit to the strong version of the Emergence
  Proposal. An even more restrictive assumption would be that  there exists a notion of exact emergence in asymptotic limits, so that
  the full moduli dependence is emerging from integrating out the  towers
of light states.

In this paper, we will shed some new light on these proposals by
generalizing the mostly single modulus computations to a model featuring
multiple moduli fields. Of course, the main obstacle in
performing a concrete computation is the precise knowledge of the light towers in the
asymptotic region one is interested in. For this reason, we consider
the simple, supersymmetric compactification of the
type IIA superstring
 on the untwisted sector of a $\mathbb Z_2\times \mathbb Z'_2$
orbifold of the 6-torus and keep track of 7 saxionic and
7 axionic moduli fields, i.e. the complex structure moduli, the
complexified K\"ahler and the complexified 4D dilaton moduli.
These are all  moduli arising in the NS-NS sector of the type IIA
superstring and for simplicity we will set all the R-R moduli to zero in our analysis.
The advantage of this model is that in the weak string coupling limit,
we know the precise moduli dependent spectrum so that we can carry out
the full loop computation of the kinetic terms keeping the full moduli dependence.
Building upon  some of the computational techniques described in
\cite{Castellano:2022bvr},
we can eventually check which components of the full
tree-level  14 dimensional   field metric emerge in this asymptotic
limit.

This paper is organized as follows:
In section 2 we define our model of interest
and make a first  approach to compute the one-loop corrections
by integrating out  only the Kaluza-Klein (KK) towers of light states.
We find that components of the field metric emerge whose
classical counterparts are vanishing. Some
of the basic notions and relations of the Emergence Proposal are
relegated to appendix \ref{app_a}.

In section 3 we first observe  that in the asymptotic weak string
coupling limit, a consistent computation
requires to include not only the KK towers but also the winding modes and the fundamental
string excitations. As we will discuss, the handling of the latter is
not straightforward, as there is an issue already with the
definition of the species scale. It turns out that a naive QFT
approach does not give exactly the same result as an approach
based on Black Holes. They differ by certain logarithmic factors
of type $\log(M_{\rm pl}/M_s)$. Such factors will accompany
our computations in the remainder of the paper.

In the second part of this section, we also  consider two other limits of type IIA on a 6-torus, namely where
one K\"ahler modulus goes to infinity and where one complex structure
modulus becomes asymptotically large. Consistent with the
Emergent String Conjecture \cite{Lee:2018urn,Lee:2019xtm,Lee:2019wij} (see also
\cite{Baume:2019sry,Lanza:2021udy}),
we will show that in these limits
there are the same number of light towers of 4D particles and
one  low-tension string as were present in the weak coupling limit.
However, now these states are mostly given by wrapped $D$-branes
and NS-branes. In fact, for the large K\"ahler modulus regime,
the relevant light 4D string is given by a wrapped NS5-brane,
whereas in the large complex structure limit it becomes
a wrapped KK-monopole. Despite the difficulty of knowing
all bound states of these branes and their masses,
in appendix \ref{app_c} we provide some admittedly speculative arguments how such
a mass formula could look like and unscrupulously use it in the body of the paper.

In section 4, we carry out a full emergence computation
including all 13 towers of light states. As we will emphasize,
only in
the QFT approach there are techniques available for such computations. As a consequence, 
ubiquitous $\log$-suppressions will appear, whose potential interpretation will be
discussed at the end of this section. 
However, it will turn out that we  recover almost the full
tree-level metric and gauge couplings.
To be more precise, no terms
are generated that were absent at tree-level even though a few singular  classical couplings are just not
generated. This  would disfavor both Variant A and Variant B
of the  Weak Emergence Proposal.
However,  the non-vanishing  ones do have the same moduli dependence
as their classical counterparts. Hence, ignoring for the moment the vanishing
singular ones, we could say that our results
  go beyond Variant A of the Weak Emergence Proposal
  and be more in favor of Variant B.
Moreover, we encounter a non-trivial  numerical
relation that combines contributions from the KK and winding towers
as well as contributions related to the string tower.
All this makes us confident  that despite the unsettled
question about the QFT versus the Black Hole picture,
our computation captures already a large portion of the story.


\section{Preliminaries}
\label{sec_emergepert}

In this section, we perform a non-trivial test of  the emergence
proposal by confronting it with a field space metric of 14
moduli. First we define more concretely the toroidal orbifold
background that we consider and attempt
to compute the one-loop corrections to the field space metric
taking only the KK-modes into account. This is meant
for  pedagogical purposes and for sharpening our computational
tools. For completeness, we have collected the basic notions and a
number of useful relations
about the Emergence Conjecture in appendix \ref{app_a}.

\subsection{The orbifold background}

\begin{figure}[ht]
    \centering
    \includegraphics[width=\textwidth]{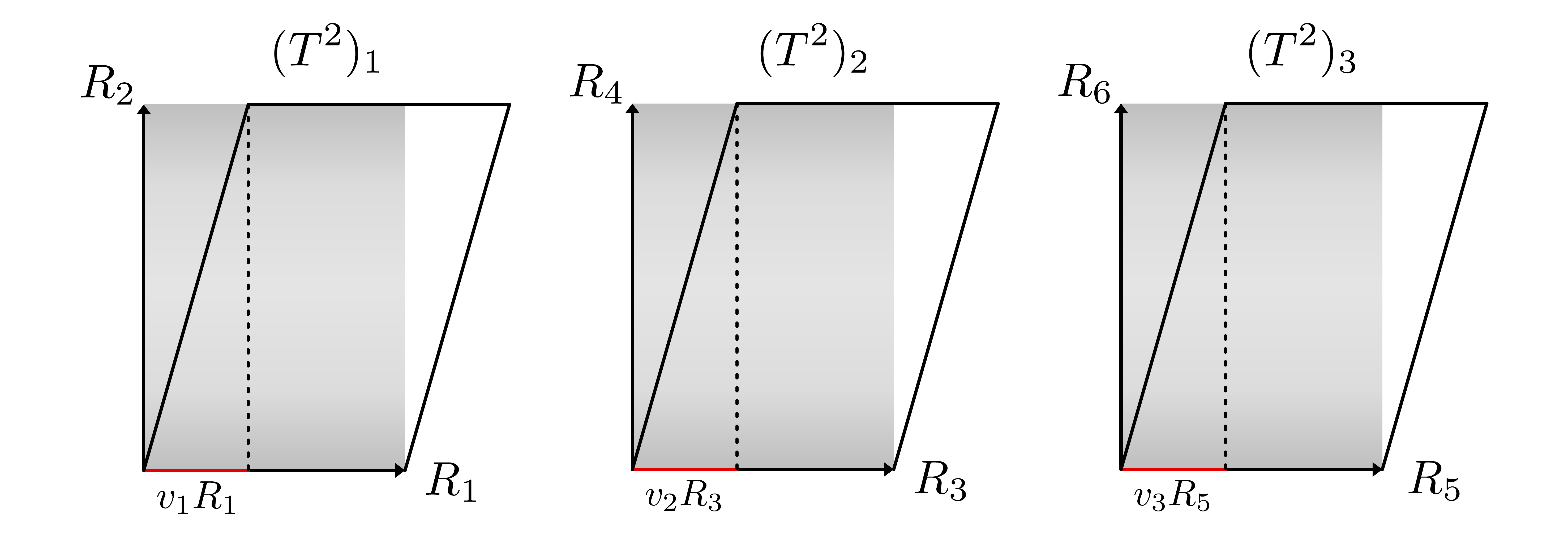}
    \caption{Conventions for the components of the factorized 6-torus. For instance, $(T^2)_1$ has a complex structure modulus with saxion $u_1 = R_2/R_1$ and the respective Kähler modulus $t_1 = R_1 R_2 / \alpha'$ is measured in string units.}
    \label{6-torus-def}
\end{figure}

We consider the type IIA superstring
 compactified on the usual $\mathbb Z_2 \times \mathbb Z'_2$
orbifold of a 6-dimensional torus.
To allow for this action, the torus
takes the factorized form $T^6=T^2\times T^2\times T^2$
so that after introducing a complex coordinate $Z_I$ on each $T^2$
the action takes the form
\eq{
       \Theta:\begin{cases}  Z_1 \to -Z_1 \\  Z_2 \to -Z_2\\
         Z_3 \to Z_3
       \end{cases}\!\!\!\!,\qquad\qquad
        \Theta':\begin{cases}  Z_1 \to Z_1 \\  Z_2 \to -Z_2\\
         Z_3 \to -Z_3
       \end{cases}\!\!\!\!.
     }
As we will see below, for our purpose we can restrict     
ourselves to the untwisted sector  of this $N=2$ supersymmetric background.
As shown in figure \ref{6-torus-def},
each $T^2$ comes with a complex structure modulus ${\cal U}_I=v_I+i u_I$ and
a complexified K\"ahler modulus ${\cal T}_I=b_I +i t_I$, where $b_I$ is
the Kalb-Ramond two-form field $B_2$ integrated over the 2-torus.
Moreover, $t_I$ is the string frame volume of the $I$-th $T^2$-factor
measured in units of $\alpha'$.
The complex structure moduli are part of three $N=2$ hypermultiplets and
the complexified K\"ahler moduli reside in three $N=2$ vector multiplets.

In order to  apply the formulas reviewed in appendix \ref{app_a},
we need to know the spectrum of particles in 4D whose mass
is below the UV cut-off scale, i.e. the species scale. To identify all
of these states is in general a non-trivial question, but in certain
asymptotic limits one might get better control. In fact, so far  the emergence
proposal has only been tested  in such  infinite field distance
limits. The best understood limit is, of course, the perturbative string
 limit, i.e. where the string coupling is very small  $g_s=e^\phi\ll 1$.
 It is appropriate to merge the  4D dilaton and its axionic partner
 into a complex field
\eq{
  S=\rho+i e^{-\phi} \sqrt{t_1 t_2 t_3} =\rho +i\sigma \,.
}
Here the axion $\rho$ is defined as the 4D magnetic dual
of the Kalb-Ramond field $B_2$ with both legs along
the 4D flat space-time. It can be thought of as the
10D magnetic dual 6-form $B_6$ with all legs along the $T^6$.
The field $S$ is the NS-NS part of an $N=2$ hypermultiplet.

In the weak coupling limit, the usual CFT string partition function
provides the lightest states, which are the KK,
winding and string oscillator modes.
The mass dependence of these states is known explicitly for
arbitrary values of the complex structure, the complexified K\"ahler moduli
and the 4D dilaton as\cite{Blumenhagen:2013fgp,Polchinski:1998rr}
\begin{eqnarray}
  \label{KKwindstrmassa}
      &&\!\!\! M^2={M^2_{\rm pl}\over \sigma^2}\bigg\{\sum_{I=1}^3 \Bigg[  \bigg( \frac{m_1^I
            -v_I m_2^I + b_I n_1^I + b_I v_I n_2^I}{  u_I^{1\over 2}\,t_I^{1\over 2}} \bigg)^2+
        \bigg( { (m_2^I - b_I n_2^I) \,   u_I^{1\over
              2} \over t_I^{1\over 2}  }\bigg)^2+\nonumber\\[0.2cm]
        &&\phantom{aaaaaaaaaaaaaa}\bigg( { (n_1^I + v_I n_2^I) \,   t_I^{1\over
              2} \over u_I^{1\over 2}  }\bigg)^2+
         \bigg( { n_2^I    u_I^{1\over
              2} \, t_I^{1\over 2}  }\bigg)^2\Bigg] +\kappa^2 N \bigg\}\,.
\end{eqnarray}
Here, the $m^I_{1,2}$ denote the  KK modes and the $n^I_{1,2}$
the winding modes\footnote{Let us note that the
    presence of these KK and winding modes are special for this
orbifold limit of a Calabi-Yau manifold, as they correspond
to torsional one-cycles which are e.g. absent for a  Calabi-Yau $X$ with
vanishing fundamental group $\pi_1(X)=0$. Strictly speaking, only
KK and winding modes are present which are invariant under
the orbifold action. However, we have convinced ourselves that
this issue only induces a change in some numerical factors.}
of the fundamental string and can be more compactly
denoted as $\vec{m}^I=(m^I_1,m^I_2)$ and $\vec{n}^I=(n^I_1,n^I_2)$\,.\footnote{$\bar{N}$ has been eliminated by imposing the level matching condition 
$N-\bar{N}+\sum_{I}\vec{m}^I\cdot\vec{n}^I=0$, which would at most change our final results by a numerical factor.}
Moreover, $N$ is the
level of the tower of string oscillator modes, which comes with a
degeneracy of states scaling at large level $N$ as
\eq{
  \label{stringdegeneracy}
                  {\rm deg}_N={\gamma\over N^{\nu\over 2} } \,   e^{\beta \sqrt{N}}\,,
}
where for later purposes  we left open the values of the parameters
$\beta,\gamma$ and $\nu$. For the 10D type IIA superstring one has e.g.
$\beta=4\pi \sqrt{2}$ and $\nu=11$.

In addition, the $\mathbb Z_2\times\mathbb Z'_2$
orbifold has 3 twisted sectors, each of them comprised of 16 fixed points of codimension 4.
Depending on whether one turns on  discrete torsion or not,
the twisted sector gives rise to 48 complex structure or K\"ahler  moduli,  respectively. 
Moreover, in each of them there are 2 KK and 2 winding modes and the
excitation number of the corresponding twisted string. We will
see in section \ref{sec_emergeKKWS} that these additive towers of light states give subdominant
contributions to the number of light species.

Observe that the axion $\rho$ does not appear in the mass formula \eqref{KKwindstrmassa}.
In the following, we will sometimes  collectively denote all 14 real
moduli 
as ${\cal M}_A$ or if we restrict to the set of 12 complex structure
and complexified K\"ahler  moduli as ${\cal M}_a$.

The 4D dilaton relates the string to the 4D Planck scale
\eq{
  M_s\simeq { M_{\rm pl}\over \sigma}\,,
}  
where we left open a numerical factor.
The classical field space metric on this 14-dimensional moduli space is known to
be of the form \cite{FERRARA1990317} (see also \cite{Grimm:2004ua,Grimm:2004uq,Baume:2019sry})
\eq{
  \label{modulimetric}
  ds^2&= g^{(0)}_{AB} \; d{\cal M}_A\,   d{\cal M}_B \\
  &={1\over \sigma^2} d\sigma^2 +{1\over 4\sigma^4} d\rho^2+\sum_{I=1}^3 {1\over 4 u_I^2} \big(du_I^2+dv_I^2\big)+\sum_{I=1}^3 {1\over 4 t_I^2}
        \big(dt_I^2+db_I^2\big)\,,
}
so that $G^{(0)}_{AB}= M_{\rm pl}^2\, g^{(0)}_{AB}$.
Note that both this metric and the mass formula enjoy a (discrete)
shift symmetry for $v_I$, $b_I$ and $\rho$ so that these moduli are (quasi)-axions.

In the type IIA case, the three complexified K\"ahler moduli are part of a full
$N=2$ vector multiplet, where the three gauge fields  come from the
dimensional reduction of the R-R three-form $C_3$ along the three 2-cycles.
There is one more vector  field, which is the graviphoton  residing in
the $N=2$ gravity multiplet. This is just the R-R one-form $C_1$.
These four 4D gauge fields are denoted as ${\cal A}^\Lambda$ (for
$\Lambda=0,1,2,3$) with field strength ${\cal F}^\Lambda =d{\cal A}^\Lambda$.

The kinetic terms   of the K\"ahler moduli and these vector fields
are governed by special geometry, which we here only very briefly summarize.
First, one defines homogeneous coordinates $X^\Lambda$ which eventually
are
related to the inhomogeneous coordinates of the K\"ahler moduli space
as $X^0=1$ and ${\cal T}_I=X^I/X^0$. Moreover, one introduces a
prepotential ${F}$ which is a homogeneous function of degree two
of  the coordinates $X^\Lambda$. Defining $F_\Lambda =\partial_\Lambda
F$, the metric on the moduli space is given in terms
of the  K\"ahler potential
\eq{
  \label{kaehlerpot}
  K=-\log\left( i\big(X^\Lambda \ov{F}_\Lambda-\ov{X}^\Lambda F_\Lambda\big) \right)\,.
 } 
The gauge kinetic terms  are then expressed  as  \cite{Bodner:1990zm}
 \eq{
  S_{\rm gauge}=-{1 \over 2} \int d^4 x \Big(
          f_{\Lambda\Sigma} \, {\cal F}^\Lambda\wedge \star {\cal  F}^\Sigma+
          \Theta_{\Lambda\Sigma} \, {\cal F}^\Lambda\wedge  {\cal F}^\Sigma \Big)\,,
        }
 where  at the classical level the gauge kinetic function
 $f_{\Lambda\Sigma}$
 and the Theta-angles $ \Theta_{\Lambda\Sigma}$ are given in  terms
 of the imaginary and real parts of the period matrix of the underlying
 Calabi-Yau threefold. The latter can be expressed in terms of the prepotential as 
 \eq{
   \label{periodmatrix}
        {\cal N}_{\Lambda\Sigma}=\ov{F}_{\Lambda\Sigma}+2i{ {\rm Im}(F_{\Lambda\Gamma})
          {\rm Im}(F_{\Sigma\Delta}) X^\Gamma X^\Delta \over
          {\rm Im}(F_{\Gamma\Delta }) X^\Gamma  X^\Delta}\, 
      }
where     $F_{\Lambda\Sigma}=\partial^2F/\partial X^\Lambda  \partial X^\Sigma$.
In our toroidal orbifold case, we have the prepotential
 \eq{
              F={X^1 X^2 X^3\over X^0}
}
leading to the  gauge couplings (see e.g. \cite{Blumenhagen:2003vr})
\eq{
  \label{gaugeterms}
    f= {\rm Im}({\cal N})=\left(\begin{matrix}  t_1 t_2 t_3 q &
             -t_2 t_3 {b_1\over t_1} &  -t_1 t_3 {b_2\over t_2} & -t_1 t_2 {b_3\over t_3} \\
             \ldots &  {t_2 t_3\over t_1} & 0  & 0 \\
             \ldots &   \ldots & {t_1 t_3\over t_2} & 0   \\
              \ldots &   \ldots &  \ldots & {t_1 t_2\over t_3}  \end{matrix}\right)
 } 
 with
 \eq{
   q=1+\left({b_1\over t_1}\right)^2+\left({b_2\over t_2}\right)^2+\left({b_3\over t_3}\right)^2\,.
}
Similarly, for  the $\Theta$-angles one gets
\eq{
  \label{Thetaterms}
   \Theta=  {\rm Re}({\cal N})=\left(\begin{matrix}  2 b_1 b_2 b_3  &
             -b_2 b_3  &  -b_1 b_3  & -b_1 b_2  \\
             \ldots &  0 & b_3  & b_2 \\
             \ldots &   \ldots &  0 & b_1   \\
              \ldots &   \ldots &  \ldots & 0  \end{matrix}\right)\,.
 }

To discriminate among the various versions of the Emergence Proposal,
the question is which parts of the tree-level moduli metric and gauge
kinetic terms can be
recovered  via integrating out  (a subset of) the
light perturbative string states at the one-loop level in the
asymptotic field regimes.

\subsection{Emergence for KK modes}
\label{Emergence_KK}

One basic ingredient is the quantum  gravity cut-off, which in the
case of a large number of light species is not the Planck-scale but
the species scale $\tilde{\Lambda}$\cite{Dvali:2007hz,Dvali:2007wp}.
In the QFT picture, one considers the quantum corrections to the graviton
propagator due to the coupling of $N_{\rm sp}$ light species to
gravity. Starting from the Einstein-Hilbert term of the action,
the species scale is defined as the mass scale 
where the one-loop contributions become of the same order as the
tree level ones. In particular, in
4D, which we will focus on in the following sections, one finds a
propagator
\eq{
  \label{graviton self energy}
    \pi^{-1}(p^2)=p^2\left(1-\frac{N_{\rm sp}\,p^2}{120\pi M^2_{\rm
          pl}}\log\left(-\frac{p^2}{\mu^2}\right) +\gamma \sum_{n=1}^{N_{\rm
          sp}}
                  {p^2\over M^2_{\rm pl}}  {m_n\over \sqrt{-p^2}}\right)\,,
}
where we also included the form of the mass dependent terms \cite{Donoghue:1994dn}
with $\gamma$ denoting  an order one parameter.

In this way,   we get the scaling\footnote{We have checked that for
    KK-towers and string towers, the mass dependent term in
    \eqref{graviton self energy} really gives $N_{\rm sp}\tilde\Lambda^2 /M_{\rm pl}^2$
    so that for the natural choice $\mu\sim \tilde \Lambda$ one indeed
    arrives at the relation \eqref{speciesscale} without any
    $\log$-correction. We thank Niccol\`o Cribiori for bringing this to
    our attention.}
\begin{equation}
  \label{speciesscale}
    \tilde{\Lambda}\sim \frac{M_{{\rm pl}}}{\sqrt{N_{\rm sp}}}\,.
\end{equation}
In practice one has two coupled relations, namely \eqref{speciesscale} and the
definition
of the number of light species
\eq{
      N_{\rm sp}=\#( m\le \tilde\Lambda )\,,  
}
which can be solved for $N_{\rm sp}$ and $\tilde\Lambda$.

In the Black Hole picture, the species scale is defined via the radius $r_0=1/\tilde\Lambda$ of the minimal-sized Black Hole that can be described within the EFT. The mass and Bekenstein-Hawking entropy of such a Black Hole are
\eq{
  \label{BekensteinH}
  M_{\rm BH}={M_{\rm pl}^2\over \tilde\Lambda}\,,\qquad\qquad 
   S_{\rm BH}={M_{\rm pl}^2\over \tilde\Lambda^2}\,.
}
The number of species is defined via  the statistical entropy as
\eq{
  \label{entropystat}
         S_{\rm BH}=\log\Omega(M_{\rm BH}) =:N_{\rm sp}\,,
}
where $\Omega(M_{\rm BH})$ is the number of ways the macroscopic
Black Hole of mass $M_{\rm BH}$ can be realized by the microstates.
Note that this definition of the number of species also 
satisfies  the relation \eqref{speciesscale}.
In practice, a second
relation between $N_{\rm sp}$ and $\tilde\Lambda$ follows from the
microcanonical relation \eqref{entropystat} so that again both are  
determined. 

\subsubsection*{Species scale for KK-modes}

In appendix \ref{app_b}, we  apply these two definitions  to a
one-dimensional KK tower of states with spacing
\eq{
    \Delta m={M_s\over r}={M_{\rm pl}\over \sigma r}\,.
  }
Here $r$ denotes the radius of the circle in units of the string length.  
In this case, the QFT approach is very simple whereas the BH approach
turns out to be a bit more involved\footnote{We acknowledge the support
  of Niccol\`o Cribiori for carrying out this computation.}.
However, at the end of the day both approaches give the same result
\eq{
            \tilde\Lambda\sim {M_{\rm pl}\over (\sigma r)^{1\over
                3}}\,,\qquad\quad
            N_{\rm sp}\sim  (\sigma r)^{2\over 3}\,.
}            
Note that the species scale is nothing else than the 5D Planck-scale.

Next, we generalize the  computation in the QFT picture to the full toroidal orbifold
model, where  we consider the 
weak string coupling limit $\sigma\to\infty$
while keeping all the other moduli in a moderate regime.
Since $\sigma$ only appears as an overall factor in the mass formula,
in the regime  $t_I>1$ the KK modes are the lightest modes.
For this truncated mass spectrum, the number of light species reads
\eq{
    N_{\rm sp}=\underbrace{\sum_{\vec{m}^I}}_{\mathclap{M_{KK}\le \tilde \Lambda^{(\rm KK)}}} 
    \approx \int \prod_{I=1}^3 dm_1^I\,dm_2^I\, ,
}
where we are summing over modes with non-zero excitations such that the total mass lies below the threshold $\tilde \Lambda^{(\rm KK)}$. In what follows, sums over excitation numbers will always be approximated by integrals, requiring a sufficiently dense spectrum to be accurate. Since the mass \eqref{KKwindstrmassa} has an overall suppression by $\sigma$, the weak coupling limit indeed justifies taking the continuum limit in all directions.  To implement the bound $M_{KK}\le \tilde \Lambda^{(\rm KK)}$, we introduce the 6 variables $x^I, y^I$ via
\eq{
      M_{\rm KK}^2={M^2_{\rm pl}\over \sigma^2} \sum_{I=1}^3 \Bigg[  &\bigg( \underbrace{{m_1^I
            -v_I m_2^I \over u_I^{1\over
              2} \,t_I^{1\over 2}}}_{x^I} \bigg)^2+
        \bigg( \underbrace{{ m_2^I \, u_I^{1\over
              2} \over t_I^{1\over 2}}}_{y^I}\bigg)^2 \Bigg] .
}
The determinant of the Jacobian of this change is ${\rm det}(J)=t_1
t_2 t_3 \equiv {\cal V}_6$. In these coordinates we are integrating
over a ball in 6 dimensions with radius $R={\tilde \Lambda^{(\rm
    KK)}\over M_{\rm pl}}\sigma$. Hence, it is convenient to introduce
the 6D spherical coordinates
\vspace{0.1cm}
\eq{
    \label{defsphere6}
    x^1&=r \cos\varphi_1 \\
    y^1&=r \sin\varphi_1\,\cos\varphi_2 \\
    x^2&=r \sin\varphi_1\,\sin\varphi_2\,\cos\varphi_3 \\
    y^2&=r \sin\varphi_1\,\sin\varphi_2\,\sin\varphi_3\,\cos\varphi_4 \\
    x^3&=r \sin\varphi_1\,\sin\varphi_2\,\sin\varphi_3\,\sin\varphi_4\,\cos\varphi_5 \\
    y^3&=r \sin\varphi_1\,\sin\varphi_2\,\sin\varphi_3\,\sin\varphi_4\,\sin\varphi_5
  }
\vspace{0.2cm}
  \noindent
with $r\ge 0$, $\varphi_1,\ldots,\varphi_{4}\in[0,\pi]$ and
$\varphi_{5}\in[0,2\pi]$. The integration measure becomes
\eq{
         \mu = dr\, r^{5}\,  d\varphi_1\ldots d\varphi_{5}  \; \sin^{4}\!\varphi_1\,
        \sin^{3}\!\varphi_2\,\sin^{2}\!\varphi_3\,  \sin\varphi_{4} \,.
}
Now, we can compute $N_{\rm sp}$ 
in the new coordinates and invoke \eqref{speciesscale} (eliminating the $\log(N_{\rm sp})$ correction by setting $\tilde{\Lambda}=\mu$ in \eqref{graviton self energy} ) to determine the species scale 
\eq{
    \tilde \Lambda^{(\rm KK)} = \frac{M_{\rm pl}}{\sigma^{3/4}} \left(
      \frac{6}{{\cal V}_6 \, {\rm vol}(S^5)}\right)^\frac{1}{8} \sim
    {\sigma^{1/4} \over {\cal V}^{1/8}_6} M_s\sim M_{\rm pl,10}  \,.
}
The quantum gravity cut-off turns out to be the ten-dimensional Planck
scale, as expected given that only the KK modes have been taken under consideration in this calculation.

\subsubsection*{One-loop corrections to the moduli space metric}

Next, let us determine the one-loop moduli metric from integrating out
the light KK modes. In appendix \ref{app_a} we recall that
these emergent one-loop diagrams mimic a classical behavior
in the sense that $\hbar$ drops out. However, in the following
we nevertheless just call them one-loop diagrams.
When we really mean the standard
stringy loop diagrams, i.e. the higher genus diagrams coming with extra
factors of the string coupling constant $g_s$, we call them
``stringy one-loop corrections''.

According to \eqref{one-loop-KK}, up to numerical prefactors the one-loop metric is given by
\eq{
  \label{metricKK}
  G^{(1)}_{{\cal M}_A {\cal M}_B}\simeq
  \underbrace{\sum_{\vec{m}^I}}_{M_{\rm KK}\le \tilde \Lambda^{(\rm KK)}}
        \!\!\! \,(\partial_{{\cal M}_A} M_{\rm KK})\, (\partial_{{\cal
            M}_B} M_{\rm KK}) \, \log\! \left( \frac{\tilde
            \Lambda^{(\rm KK)}}{M_{\rm KK}} \right)\,.
}   
To proceed, we need the  derivatives of the mass formula, which can be expressed in the compact form
\eq{
  \label{KKderiv}
    \partial_{u_I} M_{\rm KK} &\simeq {M_{\rm pl}\over \sigma} {1\over 2\, r\,u_I}\Big(-x_I^2+y_I^2\Big) =
    {M_{\rm pl}\over \sigma}{r\over 2\, u_I} \,m_{u_I}(\underline\varphi)\\[0.1cm]
    \partial_{v_I} M_{\rm KK} &\simeq {M_{\rm pl}\over \sigma} {1\over 2\, r\,u_I}\Big( -2\, x_I\, y_I \Big) ={M_{\rm pl}\over \sigma}{r\over 2\, u_I} \,m_{v_I}(\underline\varphi)\\[0.1cm]
    \partial_{t_I} M_{\rm KK} &\simeq - {M_{\rm pl}\over \sigma} {1\over 2\, r\,u_I}\Big(x_I^2 + y_I^2\Big) = {M_{\rm pl}\over \sigma}{r\over 2\, t_I} \,m_{t_I}(\underline\varphi)\\[0.1cm]
    \partial_\sigma M_{\rm KK} &\simeq -{M_{\rm pl} \over
      \sigma^2}{r}\,,\qquad  \partial_\rho M_{\rm KK}=0\,,
}
where inserting the definition of the spherical coordinates \eqref{defsphere6} allows us to determine the functions $m_{u_I}(\underline\varphi),\ldots, m_{t_I}(\underline\varphi)$ which only depend on the angular variables
$\varphi_1,\ldots,\varphi_{5}$. From the last line in \eqref{KKderiv}, we can also formally introduce  $m_{\sigma}(\underline\varphi)=1$. In the evaluation of \eqref{metricKK} the following ``angular
metric'' appears 
\eq{
    \tilde{g}_{{\cal M}_A {\cal M}_B} :=\int d\Omega_{5}\,
    m_{{\cal M}_A}(\underline\varphi)\; m_{{\cal M}_B}(\underline\varphi)\,.
}
The corresponding integrals over the 5 angular variables can be
carried out expli\-cit\-ly, yielding 
\begin{equation}
   \tilde{g}_{u_I u_I} = \tilde{g}_{v_I v_I} = \tilde{g}_{t_I t_{J\neq I}} = {\pi^3 \over 12} \,,\quad
    \tilde{g}_{t_I t_I} = {\pi^3\over 6}\,, \quad
    \tilde{g}_{\sigma t_I} = \frac{\pi^3}{3} \,,\quad \tilde{g}_{\sigma \sigma} = \pi^3 \,.
\end{equation}
The final radial integral can be evaluated explicitly using
\eq{
    \int_{0}^{r_0} r^7 \log \left( \frac{r_0}{r} \right) =\frac{r_0^8}{64} \,.
}
Collecting the remaining prefactors affecting the relative normalizations of the metric components and taking into account an overall not yet determined coefficient $\lambda$, we find 
\eq{
  \label{emergemetricKK}
    G_{u_I u_J}^{(1)} = G_{v_I v_J}^{(1)} &= \frac{1}{128} \lambda G_{u_I u_J}^{(0)} \,, \qquad G_{t_I t_I}^{(1)} =  \frac{1}{64} \lambda G_{t_I t_I}^{(0)} \,,\qquad
    G_{\sigma \sigma}^{(1)}= \frac{3}{32} \lambda G_{\sigma
      \sigma}^{(0)} \,,\\[0.1cm]
      G_{t_I \sigma}^{(1)} &= \frac{M_{\rm pl}^2}{64\, t_I \sigma}\,, \qquad G_{t_I
        t_{J \neq I}}^{(1)} = \frac{M_{\rm pl}^2}{512\, t_I t_J} \,,\qquad G_{\rho\,{\cal M}_A}^{(1)}=0 \,.
}
A comparison of this one-loop field metric  with equation
\eqref{modulimetric} reveals that the
components  involving the complex structure moduli $u_I,v_I$ have
the right relative normalization but that  all other components
are at odds. Since the mass formula does not contain the axion $\rho$,
it is immediately clear that it completely decouples. Moreover,
there are non-vanishing off-diagonal components in the second line of
\eqref{emergemetricKK} whose tree-level counterparts are zero.
In addition, the relative normalization of various diagonal components
is not consistent with the tree-level one.
Therefore, the one-loop induced metric misses some of the structure of
the tree-level one and due to the singular off-diagonal
components $G_{t_I \sigma}^{(1)}$ it is not even consistent with 
the most conservative version of Weak Emergence, namely Variant A.


\section{String towers}

\begin{figure}[ht]
    \centering
    \includegraphics[width=0.9\textwidth]{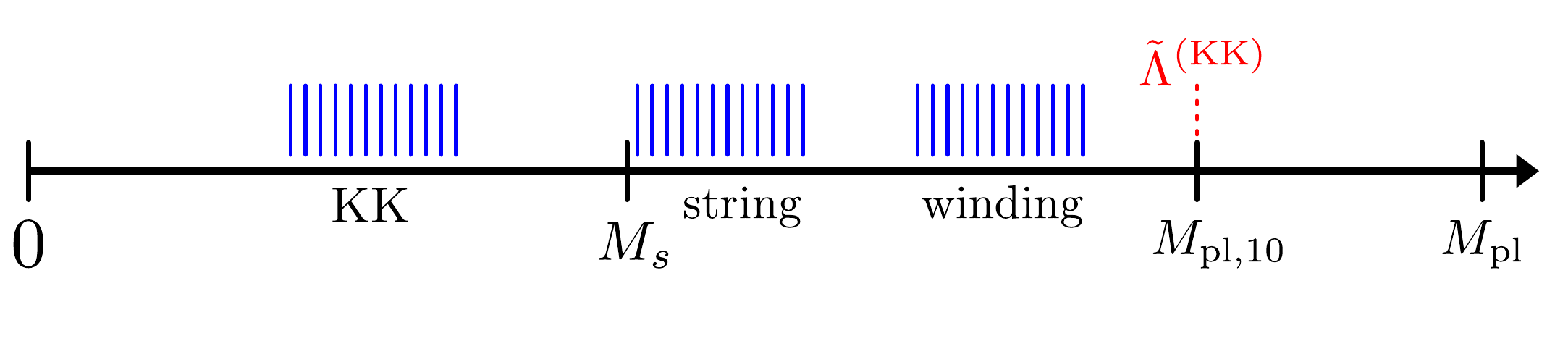}
    \caption{Tower mass scales on the energy axis. Compared to the string scale $M_s$, KK masses are suppressed by Kähler moduli, while winding modes are enhanced by them. The exponentially degenerate string excitations begin at $M_s$.}
    \label{hierarchy}
\end{figure}

The result from the previous section might not be too surprising in view of a serious short-coming of our computation, namely
that in the  large $\sigma$ limit  with all $t_I$ moderately large  one has  the hierarchy 
\eq{
    \tilde{\Lambda}^{(\rm KK)} \gg  M_{\rm wind}> M_s > M_{\rm KK}\,.
  }
 As shown in figure \ref{hierarchy},
for sufficiently large $\sigma$, both the winding modes and the string excitations turn out to be lighter than the species scale and, therefore, should have been included from the very beginning. In this case, the mass formula  treats the  complex structure
and K\"ahler moduli symmetrically and
we would have a chance to recover  the classical K\"ahler
moduli metric.

\subsection{The species scale for string towers}
\label{sec_speciesstrings}

Being forced to include the highly degenerate
tower of string excitations into  our considerations,
the first question is how to generalize the emergence computation
to this case. Unfortunately, this is not as straightforward
as one might think, as an issue already arises  for the
determination of the species scale and the number of light species.

To explain this, we have to recall the two ways to determine them, namely the QFT picture and the BH picture.
Again let us discuss this for a simplified model where one
only has a string tower with mass levels $M=M_s \sqrt{N}$ and
degeneracy ${\rm deg}_N$.
For sufficiently large mass levels $N$, one can use the asymptotic
expansion \eqref{stringdegeneracy}, which we recall here
\eq{
          {\rm deg}_N={\gamma\over N^{\nu\over 2} } \,   e^{\beta \sqrt{N}}\,.
}
Let us look at the BH picture first \cite{Dvali:2010vm}.
Then, the excitation level required for the BH mass is
\eq{
  \label{oscinumberBH}
                       \sqrt{N_{BH}}\sim {M_{\rm BH}\over M_s}\sim {M_{\rm
                         pl}^2\over M_s \tilde\Lambda_{\rm BH}}
}
which for large $\sigma$ is expected to be a very large number.
Then, up to a $\beta$ factor, the BH entropy is
\eq{
     S_{BH}\sim \log\big( {\rm deg}_{N_{BH}} \big)\sim 
     \sqrt{N_{BH}} -{\nu\over \beta} \log\big(\sqrt{N_{BH}}\big) \,.
}
Setting this equal to the Bekenstein-Hawking entropy \eqref{BekensteinH} and
using \eqref{oscinumberBH} gives an implicit equation
for $\tilde\Lambda$ which can be solved approximately by
\eq{
               \tilde\Lambda_{\rm BH}\sim M_{s} + {\nu\over \beta} M_s
               {\log\sigma\over \sigma^2} + \ldots\,.
}
Hence, for $\sigma\to\infty$ the species scale approaches the string
scale $M_s$ from above. At leading order the number of species is then
given by the entropy, i.e. 
\eq{
                       N_{\rm sp,BH}\sim {M_{\rm pl}^2\over M_s^2}\sim \sigma^2\,.
}
Note that, as mentioned in \cite{Agmon:2022thq,vandeHeisteeg:2023ubh},
this result is consistent
with the proposal \cite{vandeHeisteeg:2022btw} that for the vector multiplet moduli space, the
number of light species is given by the topological one-loop
free-energy\footnote{We thank Max Wiesner for helpful discussions on this point.}.
                     
That something is at odds with the corresponding QFT computation
can already be expected from the observation that the number of species can still be big
for large $\sigma$ with at the same time the species scale
being close to $M_s$. Indeed, 
following the same strategy as  \cite{Marchesano:2022axe,Castellano:2022bvr}
and computing the number of species by integrating over all
string states with energy  smaller than the species scale and
then solving  the resulting implicit equation \eqref{speciesscale},
one gets
\eq{
  \label{specieswithstrings}
            \tilde\Lambda_{\rm QFT} \sim M_s \log\left({M_{\rm  pl}\over
                M_s}\right)\sim M_s \log(\sigma)\,.
          }
In this case, the species scale comes out  exponentially larger than $M_s$ so
that indeed a large number of string modes
\eq{
                 N_{\rm sp,QFT}\sim {\sigma^2\over \log^2\sigma}
}
can be lighter, which
is self-consistent with having used the asymptotic expression
\eqref{stringdegeneracy}.
Let us summarize  the results in
table \ref{table_stringspecies}.

\vspace{0.3cm}
\begin{table}[ht] 
  \renewcommand{\arraystretch}{1.5} 
  \begin{center} 
    \begin{tabular}{|c|c|} 
      \hline
    QFT  picture &   BH picture        \\
      \hline \hline
      $\tilde\Lambda_{\rm QFT} \sim {M_{\rm pl}\over \sigma} \log(\sigma)$
      & $\tilde\Lambda_{\rm BH} \sim  {M_{\rm pl}\over \sigma} $ \\[0.2cm]
     $\quad N_{\rm sp}=  {\sigma^2\over \log^{2}\sigma}$ & $
           N_{\rm sp}=      {\sigma^2} \quad$ \\[0.2cm]
      \hline
      \end{tabular}
          \caption{Species scale and $N_{\rm sp}$ in QFT and BH approach.}
    \label{table_stringspecies} 
  \end{center} 
\end{table}

\noindent
Thus, even though the two computations seem to be self-consistent, they are
not mutually consistent, at least as long as  one is not willing to be
agnostic about  $\log$-factors.

Unfortunately, this tension has not been resolved yet.
However, as the authors of
\cite{Agmon:2022thq,vandeHeisteeg:2023ubh},  we tend to eventually rather
trust the Black Hole picture.
In the QFT approach we   extrapolate the quantum field theory loop
diagrams to energies that are much higher than the string scale $M_s$.
However, this high energy regime $M_s<E<M_{\rm pl}$ is strongly
believed to be very different from the sub-stringy one.
There are indications that
the string scattering amplitudes simplify in the high energy regime  \cite{Gross:1988ue}, 
and there is of course the  issue with the Hagedorn temperature $T_H\sim M_s$, where
the canonical partition functions ceases to be well defined and
a phase transition supposedly occurs \cite{Atick:1988si}.

However, the idea and the formalism of  the
Emergence Proposal are intimately linked  to the QFT approach and so far
there is no way to carry out any emergence computation using, for  instance,
the species scale and  $N_{\rm sp}$ derived from the BH picture.
Therefore, in section \ref{sec_emergstring} we will take a pragmatic approach
and explore what results we get by applying
the QFT approach for the KK, winding and string states.
Since the results for
$\tilde\Lambda$ and $N_{\rm sp}$ differ just by
some $\log$-factors, we might still see a large
portion of the emerging structure.

\subsection{Asymptotic regime \texorpdfstring{$t_1\gg 1$}{TEXT}}

Before we  move on to this computation, let us 
consider other asymptotic regions in the moduli space.
We will see that here one gets the same number of light particle
species, as well as an asymptotically tensionless string.
First consider the $t_1\gg 1$ regime.

\subsubsection*{Towers of light states}

First,  we determine  the spectrum of light states below the quantum
gravity  cut-off $\tilde\Lambda$, where for simplicity, we set all axions
to zero and consider only the dependence on the saxionic
moduli  (please see \cite{Font:2019cxq} for a related analysis).
As we will see, the mass scale of the lightest modes is
$M_{\rm pl}/\sqrt{t_1}$. 

\paragraph{Perturbative string states:}
From the KK and string winding modes,  the KK-modes on the first torus are 
the lightest ones with masses
\eq{\label{largetKKs}
          M^{(1)}_{\rm KK}\sim {M_{\rm pl}\over \sqrt{t_1}}
              {1\over \sigma \sqrt{u_1}}\,,\qquad
              M^{(2)}_{\rm KK}\sim {M_{\rm pl}\over \sqrt{t_1}}
               {\sqrt{u_1}\over \sigma}\,.
             }
All the other 4 KK- and 6 string winding modes are much heavier.

\paragraph{Wrapped $D$-brane states:}             
As already pointed out in \cite{Grimm:2018ohb}, in this limit one needs to take
into account wrapped $D$-brane states, as well. As long as
these $D$-branes do not wrap any cycle on the first $T^2$ factor,
their masses also scale like $M_{\rm pl}/\sqrt{t_1}$. 
The resulting light modes are listed in table
\ref{table_Dbranes}.
\begin{table}[H] 
  \renewcommand{\arraystretch}{1.5} 
  \begin{center} 
    \begin{tabular}{|c|c|c|} 
      \hline
    D-branes &   wrapping &  mass scale       \\
      \hline \hline
    $D0$  &  $(--,--,--)$  &  $M^{(3)}_{D0}\sim {M_{\rm pl}\over \sqrt{t_1}}
                             {1\over \sqrt{t_2 t_3}}$ \\
      \hline
       $D2$  &  $(--,++,--)$  &  $M^{(4)}_{D2}\sim {M_{\rm pl}\over \sqrt{t_1}}
                                \sqrt{t_2\over t_3}$ \\
          &  $(--,--,++)$  &  $M^{(5)}_{D2}\sim {M_{\rm pl}\over \sqrt{t_1}}
                                 \sqrt{t_3\over t_2}$ \\
           &  $(--,+-,+-)$  &  $M^{(6)}_{D2}\sim {M_{\rm pl}\over \sqrt{t_1}}
                                  {1\over \sqrt{u_2u_3}}$ \\
           &  $(--,-+,-+)$  &  $M^{(7)}_{D2}\sim {M_{\rm pl}\over \sqrt{t_1}}
                                  \sqrt{u_2u_3}$ \\
         &  $(--,+-,-+)$  &  $M^{(8)}_{D2}\sim {M_{\rm pl}\over \sqrt{t_1}}
                                \sqrt{u_3\over u_2}$ \\
          &  $(--,-+,+-)$  &  $M^{(9)}_{D2}\sim {M_{\rm pl}\over \sqrt{t_1}}
                             \sqrt{u_2\over u_3}$ \\
      \hline
       $D4$  &  $(--,++,++)$  &  $M^{(10)}_{D4}\sim {M_{\rm pl}\over \sqrt{t_1}}
                             \sqrt{t_2 t_3}$ \\
      \hline
      \end{tabular}
          \caption{Light wrapped $D$-brane states. In the second column
            we indicate by a $+$ which cycles on the internal $T^2\times
            T^2\times T^2$ are wrapped by the branes.}
    \label{table_Dbranes} 
  \end{center} 
\end{table}
\noindent
Note that all $D$-brane masses do not contain any factor of $\sigma$.

\paragraph{Wrapped NS5-brane states:}   
It turns out that the wrapped $NS5$-brane contributes two more light modes,
which are listed in table
\ref{table_NSbranes}.
\begin{table}[ht] 
  \renewcommand{\arraystretch}{1.5} 
  \begin{center} 
    \begin{tabular}{|c|c|c|} 
      \hline
    Brane &   wrapping &  mass scale       \\
      \hline \hline
    $NS5$  &  $(+-,++,++)$  &  $M^{(11)}_{NS5}\sim {M_{\rm pl}\over \sqrt{t_1}}
                              {\sigma \over \sqrt{u_1}}$\\
       &  $(-+,++,++)$  &  $M^{(12)}_{NS5}\sim {M_{\rm pl}\over \sqrt{t_1}}
                              {\sigma \sqrt{u_1}}$\\
  \hline
      \end{tabular}
          \caption{Light wrapped $NS5$-brane states.}
     \label{table_NSbranes}  
  \end{center} 
\end{table}
Hence, in total we have found 12 light modes which is precisely
the same number as found in the perturbative string limit
in section \ref{sec_emergepert}. Moreover, upon exchanging $\sigma\leftrightarrow
\sqrt{t_1}$ and $u_2\leftrightarrow t_3$ their masses follow the same pattern.
This analogy suggests that there might also be a low-tension string
in 4D, which in the former perturbative limit was just the fundamental
string.

\paragraph{Light 4D string:}   

First, we observe that e.g. a wrapped $D2$-brane  yields a string tension
that scales like
\eq{
                  T_{D2}\sim {M_{\rm pl}^2\over \sqrt{t_1}}\, ,
}
whose excitations provide 4D particles of mass
$m\sim \sqrt{T_{D2}}\sim  {M_{\rm pl}/ t^{1/4}_1}$. Thus, these
modes are parametrically heavier than the light modes listed so far.
However, we can also wrap the $NS5$-brane on the last two $T^2$ factors
yielding a string with tension and corresponding excitations
\eq{
  T_{NS5}= {M^2_{\rm pl}\over t_1}\ \Rightarrow\ 
   M^{(13)}_{NS5}= {M_{\rm pl}\over \sqrt{t_1}}\,.
 }
We think it is compelling that also in the asymptotic regime $t_1\to
\infty$,
one finds the same pattern of 12 light particles and one low-tension
string in 4D as we have seen in the much better understood
perturbative limit $\sigma\to\infty$. Additionally, the appearance of an emergent string due to an $NS5$ brane in this limit is also in agreement with \cite{Lee:2019wij,Marchesano:2022axe,Castellano:2022bvr}.

\vspace{0.2cm}
One might wonder whether, if there is an asymptotically tensionless
  string in 4D, maybe there also exists an asymptotically tensionless membrane.
  However, in accordance with the Emergent String
  Conjecture \cite{Lee:2018urn,Lee:2019xtm,Lee:2019wij},
  wrapping $D$-branes and NS5-branes such that one gets a membrane $M_2$
  in  4D, leads to its excitations of mass $M\sim T_{M_2}^{1/3}$ always being heavier than
  $M_{\rm pl}/\sqrt{t_1}$.

\subsection{Asymptotic regime \texorpdfstring{$u_1\gg 1$}{TEXT}}

For completeness, let us now consider the type IIA superstring compactified
on $T^6=(T^2)^3$ in the
limit $u_1\to\infty$. Again, we set all axions
to zero.

\paragraph{Perturbative string states:}
One  KK- and one winding mode on the first torus will be lighter than
the others
\eq{
          M^{(1)}_{\rm KK}\sim {M_{\rm pl}\over \sqrt{u_1}}
              {1\over \sigma \sqrt{t_1}}\,,\qquad
              M^{(2)}_{\rm wind}\sim {M_{\rm pl}\over \sqrt{u_1}}
               {\sqrt{t_1}\over \sigma}\,.
             }

\paragraph{Wrapped $D$-brane states:}
The light $D$-branes are the ones that wrap the $x$-cycle on the first
$T^2$ and some other directions on the second and third $T^2$.
As a consequence, only the $D2$ and $D4$ branes do lead
to light modes scaling as $M_{\rm pl}/\sqrt{u_1}$.
They are listed in table \ref{table_DbraneUs}.
\begin{table}[H] 
  \renewcommand{\arraystretch}{1.5} 
  \begin{center} 
    \begin{tabular}{|c|c|c|} 
      \hline
    $D$-branes &   wrapping &  mass scale       \\
      \hline \hline
       $D2$  &  $(+-,+-,--)$  &  $M^{(3)}_{D2}\sim {M_{\rm pl}\over \sqrt{u_1}}
                                {1\over \sqrt{u_2 t_3}}$ \\
          &  $(+-,-+,--)$  &  $M^{(4)}_{D2}\sim {M_{\rm pl}\over \sqrt{u_1}}
                                 \sqrt{u_2\over t_3}$ \\
           &  $(+-,--,+-)$  &  $M^{(5)}_{D2}\sim {M_{\rm pl}\over \sqrt{u_1}}
                                  {1\over \sqrt{t_2 u_3}}$ \\
           &  $(+-,--,-+)$  &  $M^{(6)}_{D2}\sim {M_{\rm pl}\over \sqrt{u_1}}
                                  \sqrt{u_3\over t_2}$ \\
      \hline
       $D4$  &  $(+-,-+,++)$  &  $M^{(7)}_{D4}\sim {M_{\rm pl}\over \sqrt{u_1}}
                                \sqrt{u_2 t_3}$ \\
         &  $(+-,+-,++)$  &  $M^{(8)}_{D4}\sim {M_{\rm pl}\over \sqrt{u_1}}
                                 \sqrt{t_3\over u_2}$ \\
         &  $(+-,++,-+)$  &  $M^{(9)}_{D4}\sim {M_{\rm pl}\over \sqrt{u_1}}
                                \sqrt{t_2 u_3}$ \\
         &  $(+-,++,+-)$  &  $M^{(10)}_{D4}\sim {M_{\rm pl}\over \sqrt{u_1}}
                                \sqrt{t_2\over u_3}$ \\
      \hline
      \end{tabular}
          \caption{Light wrapped $D$-brane states.}
    \label{table_DbraneUs} 
  \end{center} 
\end{table}
\vspace{-7.5mm}
A closer look reveals that, so far,  the spectrum of light states is related
to the one in the $t_1\to\infty$ limit by a T-duality in two
directions,
e.g. exchanging $u_1\leftrightarrow t_1$ and $u_3\leftrightarrow t_3$.

\paragraph{Wrapped $NS5$-brane states:}   
It turns out that there is only one light state coming from a  wrapped
$NS5$-brane, namely the 5-brane wrapping the $x$-direction on the
first $T^2$ and both directions in the remaining $T^2$'s. This provides the eleventh light mode so that relative
to the former asymptotic regimes, one state seems to be missing.
However, by applying the just mentioned T-duality transformation,
we realize that the former $NS5$-brane is mapped to a KK-monopole (see e.g. \cite{Sorkin:1983ns,Gross:1983hb,Plauschinn:2018wbo}).
Hence, we find the two additional $NS$-branes  listed in
table \ref{table_$NS$branesU}.

\begin{table}[ht] 
  \renewcommand{\arraystretch}{1.5} 
  \begin{center} 
    \begin{tabular}{|c|c|c|} 
      \hline
    Brane &   wrapping &  mass scale       \\
      \hline \hline
    $NS5$  &  $(+-,++,++)$  &  $M^{(11)}_{NS5}\sim {M_{\rm pl}\over \sqrt{u_1}}
                              {\sigma \over \sqrt{t_1}}$\\
   KK-monopole    &  $(-+,++,++)$  &  $M^{(12)}_{\rm KK-monop.}\sim {M_{\rm pl}\over \sqrt{u_1}}
                              {\sigma \sqrt{t_1}}$\\
  \hline
      \end{tabular}
          \caption{Light wrapped $NS$-brane states.}
    \label{table_$NS$branesU} 
  \end{center} 
\end{table}
We notice that when expressed in string units, the contribution from the wrapped KK-monopole is
\eq{
      M= {M_{\rm pl}\over \sqrt{u_1}}
                              {\sigma \sqrt{t_1}}={M_s\over g_s^2} r_1^2\, r_2\, r_3\, r_4\, r_5\, r_6\,.
}
The, at first sight strange, factor $r_1^2$ comes from the energy $f^2$
of the non-trivial $U(1)$ field strength $f=da$ supported on the KK-monopole.

\paragraph{Light 4D string:}   

Now it is clear that the lightest 4D string also arises from the
KK-monopole wrapped on the last two $T^2$ factors.
This yields a tension and the corresponding  masses are
\eq{
  T_{\rm KK-monop.}={M^2_s\over g_s^2} r_1^2 \, r_3\, r_4\, r_5\, r_6
  = {M^2_{\rm pl}\over u_1}\ \Rightarrow\ 
   M^{(13)}_{\rm KK-monop.}= {M_{\rm pl}\over \sqrt{u_1}}\,.
 }
 Hence, again,  in this asymptotic region of a large complex structure
we have identified precisely 12 light
particles and one low-tension string.


\section{Emergence in asymptotic regions}
\label{sec_emergstring}

Being aware of the potential limitations, in this  section we carry out a complete emergence
computation in the only currently accessible QFT approach.
Beyond the KK modes, we will  include the  winding and string modes.
First, we do this  for the perturbative string limit ($\sigma\to\infty$)
and afterwards also consider  the generalization to the limits
of a single large K\"ahler modulus and a single large complex structure modulus.
Finally, we discuss the implications of our concrete computation
for the Emergence Proposal.

\subsection{Emergence in the  weak string coupling limit}
\label{sec_emergeKKWS}

Employing similar computational methods as for the already presented KK  example, let
us now integrate out all light towers of states with a mass smaller
than the species scale.

\subsubsection*{The species scale}

For the latter, we first need to compute $N_{\rm sp}$ as given by \eqref{Nsp general}
\eq{
        N_{\rm sp}=\underbrace{\sum_{\vec{m}^I,\vec{n}^I,N}}_{M\le \tilde\Lambda}
        \!\!\!  {\rm deg}_N \approx \int \prod_{I=1}^3 dm_1^I\, dm^I_2 \,dn_1^I \, dn^I_2\, dN\; {\rm deg}_N\, ,
      }
where in the $\sigma\gg 1$ regime we can again safely approximate the sum by
an integral. This time, we need to define 13 new variables $w^I,x^I,y^I,z^I$ and $q$ via
\begin{eqnarray}
  \label{masstower}
      &&M^2={M^2_{\rm pl}\over \sigma^2}\Bigg\{ \sum_{I=1}^3 \Bigg[  \bigg( \overbrace{{m_1^I
            -v_I m_2^I + b_I n_1^I + b_I v_I n_2^I\over  u_I^{1\over
              2} \,t_I^{1\over 2}  }}^{w^I}\bigg)^2+
        \bigg(\overbrace{ { (m_2^I - b_I n_2^I) \,   u_I^{1\over
              2} \over t_I^{1\over 2}  }}^{x^I}\bigg)^2+ \nonumber \\[0.2cm] 
        &&\phantom{aaaaaaaaaaaaaa}\bigg(\underbrace{ { (n_1^I + v_I n_2^I) \,   t_I^{1\over
              2} \over u_I^{1\over 2} }}_{y^I} \bigg)^2+
         \bigg(\underbrace{ { n_2^I    u_I^{1\over
              2} \, t_I^{1\over 2}  }}_{z^I}\bigg)^2\Bigg] +\underbrace{\kappa^2 N}_{q^2}\Bigg\} \,.
\end{eqnarray}
The determinant of the Jacobian for this change of variables is $\det(J)=2 q/\kappa^2$.
In these variables, we now integrate over a ball in 13 dimensions with radius
$R={\tilde\Lambda\over M_{\rm pl}}\sigma$, which makes it convenient to
introduce  13D spherical coordinates
\eq{
  \label{defsphere13}
      q&=r \cos\varphi_0 \\
       w^1&=r \sin\varphi_0\,\cos\varphi_1 \\
       x^1&=r \sin\varphi_0\,\sin\varphi_1\,\cos\varphi_2 \\
       y^1&=r \sin\varphi_0\,\sin\varphi_1\,\sin\varphi_2\,\cos\varphi_3 \\
              &\qquad\qquad\vdots\\
       y^3&=r \sin\varphi_0\,\sin\varphi_1\ldots\ldots\ldots   \sin\varphi_{10}\, \cos\varphi_{11} \\
      z^3&=r \sin\varphi_0\,\sin\varphi_1\ldots\ldots\ldots \sin\varphi_{10} \,\sin\varphi_{11} 
}
with $r\ge 0$, $\varphi_0,\ldots,\varphi_{10}\in[0,\pi]$ and
$\varphi_{11}\in[0,2\pi]$. The measure then becomes
\eq{
        \mu=     dr\, r^{12}  d\varphi_{0} \,\sin^{11}\!\varphi_0 \,
              d\varphi_1\ldots d\varphi_{11}  \; \sin^{10}\!\varphi_1\,
            \sin^{9}\!\varphi_2\ldots   \sin\varphi_{10}      \,.
}
Putting everything together, we arrive at the following  integral for $N_{\rm sp}$
\eq{
        N_{\rm sp}\simeq{2\gamma \kappa^{\nu-2}}\, {\rm vol}(S^{11}) \int_0^{\tilde\Lambda/M_s} dr\,
        r^{13-\nu} \int_0^{\pi/2} d\varphi_{0} \,{\sin^{11}\varphi_0\over
          \cos^{\nu-1}\varphi_0}
        \exp\Big( {\textstyle {\beta\over \kappa}} r\cos\varphi_0\Big)\, ,
}     
where we have carried out already the integral over the angular
variables $\varphi_1,\ldots,\varphi_{11}$, which gives the volume ${\rm
  vol}(S^{11}) =\pi^6/60$ of the unit sphere $S^{11}$.
The large $\tilde\Lambda/M_s$ behavior of the $r$ and $\varphi_0$
integral can be determined analytically and reads
\eq{
  \label{phizerointa}
  \int_0^{r_0} dr\, r^k \int_0^{\arccos\left(\kappa/r\right)} \!\! d\varphi_0\,
       {\sin^{2n+1}\varphi_0\over \cos^m \varphi_0} \,e^{ \alpha r \cos\varphi_0}
       = {\delta_n\over \alpha^{n+2}}\, r_0^{k-n-1}\, e^{\alpha r_0}+\ldots\,,
}
with the values of the coefficients $\delta_n$ listed
in table \ref{table_deltas} for the first values of $n$.
\begin{table}[ht] 
  \setlength{\tabcolsep}{10pt}
  \renewcommand{\arraystretch}{1.2} 
  \begin{center}
    \begin{tabular}{|c||c|c|c|c|c|c|c|c|} 
    \hline
    $n$ & 0 & 1 & 2 & 3 & 4 & 5 & 6 & 7 \\ [0.5ex] 
    \hline 
    $\delta_n$ & 1 & 2 & 8 & 48 & 384 & 3840 & 46080 & 645120 \\ 
    \hline 
    \end{tabular}
    \caption{Values of $\delta_n$ coefficients.}
    \label{table_deltas} 
  \end{center} 
\end{table}
For the convergence of the integral it was important that we imposed the
physical constraint $N\ge 1$, which modifies the upper bound of the
$\varphi_0$ integration accordingly.
It is easy to see that the
integral \eqref{phizerointa} is independent of the exponent $m$ of the
$\cos\varphi$-factor
in the denominator. Applying this relation for $n=5$, $k=13-\nu$, $r_0=\tilde\Lambda/M_s$
and $\alpha=\beta/\kappa$  leads to the number of light species
\eq{
        N_{\rm sp}\simeq  7680 {\gamma \kappa^{\nu+5}\over \beta^7}  {\rm
          vol}(S^{11})
        \left({M_s\over \tilde\Lambda}\right)^{\nu-7}
        e^{{\beta\over \kappa} {\tilde\Lambda\over M_s}}\,.
      }
Using  \eqref{speciesscale}, one gets a
transcendental equation for the species scale $\tilde\Lambda$
\eq{
  \label{defeqlambda}
                {M_{\rm pl}^2\over M_s^2} = \underbrace{7680 {\gamma \kappa^{\nu+5}\over \beta^7}  {\rm
          vol}(S^{11})}_{A}
        \left({\tilde\Lambda\over M_s}\right)^{9-\nu}
        e^{{\beta\over \kappa} {\tilde\Lambda\over M_s}}\,.
      }
This is solved by the Lambert function $W(y)$\footnote{The Lambert function $W(y)$ is defined as the solution to the equation
  $x e^x=y$, for $y\ge -e^{-1}$ and
  is a multivalued function when $-e^{-1}<y<0$. The branch satisfying $-1\leq W(y)$ is called the principle branch and is denoted as $W_0(x)$, while the one satisfying $W(y)\leq -1$ is denoted by $W_{-1}(x)$. If $y\ge 0$, then $W(y)=W_0(y)$.}.
  For $\nu\le 9$ one chooses the branch $W_0$ and for $\nu> 9$ the branch
$W_{-1}$.
\eq{
        {\tilde\Lambda\over M_s}=(9-\nu) {\kappa\over \beta}\;
        W\!\left({1\over (9-\nu)} {\beta\over \kappa A^{1\over 9-\nu}}   \left({M_{\rm
                pl}\over M_s}\right)^{2\over 9-\nu}\right)\,.
} In any case, in the asymptotic regime $\sigma=M_{\rm pl}/M_s\gg 1$ one gets\footnote{
    For us, the following expansions are relevant \cite{Corless1996OnTL}:
\begin{equation*}
\begin{split}
    W_0(x\rightarrow\infty)=&\log(x)-\log(\log(x))+\ldots\\
    W_{-1}(x\rightarrow 0)=&\log(-x)-\log(-\log(-x))+\ldots
    \end{split}
\end{equation*}}
\eq{
  \label{finalspecies}
     {\tilde\Lambda\over M_s}\sim {2\kappa\over \beta}\log\left({M_{\rm
                pl}\over M_s}\right)\,,
        }
which is also independent of $\nu$. Hence, as already
mentioned in section \ref{sec_speciesstrings},  namely in equation \eqref{specieswithstrings},
after including all the light modes in the perturbative string regime
the actual UV cut-off is essentially the string scale however amplified
by a logarithmic correction.

Let us comment on the contribution to  $N_{\rm sp}$ from the 48 twisted
sectors of the orbifold. As mentioned, these provide additive towers
of light states with in each case $\Delta_{\rm t}=4$ KK and winding
modes and the fundamental twisted string. Going through the same
steps as in the untwisted case, one realizes that their number of light
states is suppressed relative to the contribution from the untwisted
sector with its $\Delta_{\rm u}=12$ KK and winding modes 
\eq{
                              {N_{\rm sp,u}\over N_{\rm sp,t}} \sim
                              \left( {\tilde\Lambda\over
                                  M_s}\right)^{(\Delta_{\rm
                                  u}-\Delta_{\rm t})\over 2}\gg 1 \,.  
}
Thus, they can be safely neglected in the asymptotic limit.

\subsubsection*{One-loop corrections to the moduli space metric}

First, we compute the one-loop correction to
$G^{(0)}_{\sigma\sigma}$. Again, we only keep the numerical factors affecting the relative normalization of the metric components. Since $\sigma$ appears in the mass formula
\eqref{masstower} only as a prefactor, the relation
\eqref{one-loop-string} becomes
\eq{
  G^{(1)}_{\sigma\sigma}\simeq
  \underbrace{\sum_{\vec{m}^I,\vec{n}^I,N}}_{M\le \tilde\Lambda}
        \!\!\!  {\rm deg}_N   \left({M_{\rm pl}\, r\over
            \sigma^2}\right)^2 \,.
      }
We can proceed in the same manner as for the computation of
$N_{\rm sp}$ and  arrive at\footnote{We have checked that using the
  exact form of the integral \eqref{wf_renorm_fermion_inti} gives the
  same   result in  the asymptotic limit   $\tilde\Lambda/M_s\gg 1$.}
\eq{
       G^{(1)}_{\sigma\sigma}\simeq {M_{\rm pl}^2\over \sigma^4} \underbrace{7680 {\gamma \kappa^{\nu+5}\over \beta^7}  {\rm
          vol}(S^{11})}_{A}
        \left({\tilde\Lambda\over M_s}\right)^{9-\nu}
        e^{{\beta\over \kappa} {\tilde\Lambda\over M_s}}\,.
      }
Using the relation \eqref{defeqlambda}, we  bring this to the
simple form
\eq{
         G^{(1)}_{\sigma\sigma}\simeq {M_{\rm pl}^2\over \sigma^2}\,.
}
This is precisely the tree-level metric $G^{(0)}_{\sigma\sigma}$ from
\eqref{modulimetric}. In particular, we realize that the still left
open coefficients $\beta,\gamma,\kappa,\nu$ completely drop out in the
final result.
Before getting too excited, we have to keep in
mind that this result is only true up to some overall numerical prefactor.

Next, we determine the one-loop corrections to all the other metric
components, again using \eqref{one-loop-string}
\eq{
  \label{metricall}
  G^{(1)}_{{\cal M}_a {\cal M}_b}\simeq
  \underbrace{\sum_{\vec{m}^I,\vec{n}^I,N}}_{M\le \tilde\Lambda}
        \!\!\!  {\rm deg}_N \,  (\partial_{{\cal M}_a} M)\, (\partial_{{\cal M}_b} M)\,.
}      
Proceeding analogously to our previous calculation for the KK
spectrum, the derivatives now take the more symmetric form

\eq{
  \label{allderiv}
\partial_{u_I} M &\simeq {M_{\rm pl}\over \sigma} {1\over 2\, r\,
  u_I}\Big( -w_I^2+x_I^2-y_I^2 +z_I^2\Big) ={M_{\rm pl}\over \sigma}
{r\sin\varphi_0\over 2\, u_I} \,m_{u_I}(\underline\varphi)\\[0.1cm]
\partial_{v_I} M &\simeq {M_{\rm pl}\over \sigma} {1\over 2\, r\,
  u_I}\Big( -2\, w_I\, x_I  +2\, y_I\, z_I\Big) ={M_{\rm pl}\over \sigma}
{r\sin\varphi_0\over 2\, u_I} \,m_{v_I}(\underline\varphi)\\[0.1cm]
\partial_{t_I} M &\simeq {M_{\rm pl}\over \sigma} {1\over 2\, r\,
  u_I}\Big( -w_I^2-x_I^2+y_I^2 +z_I^2\Big) ={M_{\rm pl}\over \sigma}
{r\sin\varphi_0\over 2\, t_I} \,m_{t_I}(\underline\varphi)\\[0.1cm]
\partial_{b_I} M &\simeq {M_{\rm pl}\over \sigma} {1\over 2\, r\,
  u_I}\Big( 2\, w_I\, y_I  -2\, x_I\, z_I\Big) ={M_{\rm pl}\over \sigma}
{r\sin\varphi_0\over 2\, t_I} \, m_{b_I}(\underline\varphi)\\[0.1cm]
\partial_\sigma M &\simeq -{M_{\rm pl} \over \sigma^2}{r} \,,
}
where inserting the definition of the spherical coordinates \eqref{defsphere13} allows
us to determine the functions $m_{u_I}(\underline\varphi),\ldots,  m_{b_I}(\underline\varphi)$
which only depend on the angular variables
$\varphi_1,\ldots,\varphi_{11}$.
Again, we can also formally introduce
$m_{\sigma}(\underline\varphi)=1$ and
in the evaluation of \eqref{metricall} an angular metric  appears 
\eq{
  \tilde{g}_{{\cal M}_A {\cal M}_B} :=\int d\Omega_{11}\,
      m_{{\cal M}_A}(\underline\varphi)\; m_{{\cal M}_B}(\underline\varphi)\,.
}
The corresponding integrals over the 11 angular variables can be
carried out explicitly which yields the only non-vanishing components
\eq{
 \tilde{g}_{u_I u_I} = \tilde{g}_{v_I v_I}={\pi^6\over 1260} \,,\qquad
  \tilde{g}_{t_I t_I}=\tilde{g}_{b_I b_I}={\pi^6\over 1260}\,,
 }
 so that
 \eq{
   \tilde{g}_{{\cal M}_a {\cal M}_b}={\pi^6\over 1260}\delta_{{\cal M}_a {\cal
       M}_b}\,.
   }
All off-diagonal components including
$\tilde{g}_{u_I \sigma},\ldots,\tilde{g}_{b_I \sigma}$ vanish. Hence,
from these integrals we already get a large portion of the structure of
the tree-level metric $G^{(0)}_{AB}$.
It remains to carry out the full integrals, i.e. also the ones over $r$ and $\varphi_0$
\eq{
        G^{(1)}_{{\cal M}_a {\cal M}_a} \simeq M_{\rm pl}^2\, &{2\gamma \kappa^{\nu-2}\over 4
          {{\cal M}^2_a}\sigma^2} \,\tilde{g}_{{\cal M}_a {\cal M}_a} \!\\
        &\int_0^{\tilde\Lambda/M_s}\!\!\! dr\,
        r^{15-\nu} \int_0^{\arccos\left(\kappa/r\right)} \!\! d\varphi_{0} \,{\sin^{15}\varphi_0\over
          \cos^{\nu-1}\varphi_0}
        \exp\!\Big( {\textstyle {\beta\over \kappa}} r\cos\varphi_0\!\Big)\,.
      }
Next, employing \eqref{phizerointa}  for $n=7$, $k=15-\nu$, $r_0=\tilde\Lambda/M_s$
and $\alpha=\beta/\kappa$ and using  the relation  \eqref{defeqlambda}
one  arrives at
\eq{
  \label{metricprevious}
        G^{(1)}_{{\cal M}_a {\cal M}_a} \simeq M_{\rm pl}^2\, {\kappa^2\over \beta^2}
           {168\, \tilde{g}_{{\cal M}_a {\cal M}_a}\over {\rm
            vol}(S^{11})}{1\over 4 {\cal M}_a^2} \left({M_s\over \tilde\Lambda}\right)^2\, ,
      }
      where we used $\delta_7/\delta_5=168$.
First, we observe the intriguing numerical relation
\eq{
  \label{wonderrel}
                       {168\over 8\, {\rm
                           vol}(S^{11})}  \, \tilde{g}_{{\cal M}_a {\cal M}_a}=
                   {168\cdot 60 \over 8\, \pi^6}   {\pi^6\over 1260}=1\,,
                 }
where the factors ${\rm vol}(S^{11})$ and  $\tilde{g}_{{\cal M}_a {\cal
 M}_a}$ were coming from integration over KK and
winding modes and the factor $168$ from the integration over
the string oscillators, namely the  integral \eqref{phizerointa}.
Hence, it seems that something highly non-trivial
between extra dimensions and strings is happening here,
in fact on a quantitative level.
      
Then, invoking in addition  the species scale \eqref{finalspecies},
we  finally get the one-loop moduli metric
\eq{
  \label{UTmetric}
        G^{(1)}_{{\cal M}_a {\cal M}_b} \simeq {M_{\rm pl}^2\over 2\, {\cal M}_a^2}\,
        {1\over \log^2\!\big({M_{\rm pl}\over M_s}\big)} \,\delta_{{\cal M}_a {\cal M}_b}\,.
      }
At this point it is tempting to speculate that working with the
species scale found in the BH picture, namely $\tilde\Lambda\sim M_s$,
the $\log^2$-factor would be absent. However, the previous relation
\eqref{metricprevious} was of course derived in the QFT picture.

Taking into account an overall not yet determined coefficient $\lambda$ and that the axion $\rho$ completely decoupled, we can
summarize the results for the one-loop field metric as
\eq{
            \label{metric_full_string}
            &G^{(1)}_{\sigma \sigma}={\lambda}\,
            G^{(0)}_{\sigma\sigma}\,,\qquad  G^{(1)}_{\rho
              \rho}=0\,, \qquad  G^{(1)}_{\rho
              \sigma}=0\,, \\[0.1cm]
       &G^{(1)}_{\sigma {\cal
                M}_a}=0\,, \qquad\quad\, G^{(1)}_{\rho {\cal
                M}_a}=0\,,\\[0.1cm]
            \qquad
             &G^{(1)}_{{\cal M}_a {\cal M}_b}= {2 \lambda\over
              \log^2 (\sigma )}\,G^{(0)}_{{\cal M}_a {\cal M}_b}
          \,.
        }
Thus, a large portion of the structure of the tree-level metric
is there, in particular all off-diagonal components of the metric
are vanishing. However, the classical singular behavior of $G_{\rho
  \rho}$ is not reproduced, which if taken seriously threatens both Variant A and
Variant B of the  Weak Emergence Proposal.
On the positive  side, up to the $\log(\sigma)$ factor,
  the dependence on all  orthogonal moduli is reproduced correctly, in agreement with Variant B of the Weak Emergence Proposal. However, the $\log$-factors that we have seen already in the
result of the species scale also make their appearance for the
one-loop field metric.

\subsubsection*{One-loop corrections to the gauge kinetic terms}

We can  also compute the one-loop corrections to the gauge kinetic
terms. However, since all perturbative string states are neutral with
respect to the R-R gauge fields, it is immediately clear that
integrating them out does not lead to any contribution so that
\eq{
    f^{(1)}_{IJ}=0\,.
 }
The charged states are given by $D0$ and wrapped $D2$ branes, which in
the asymptotic region $\sigma\gg 1$ are heavier than the species scale.
Hence, the gauge kinetic terms cannot  emerge in this limit.
Note that  the tree-level
   ones are singular in the K\"ahler moduli and not in $\sigma$, which
   is taken to infinity here. Therefore, we can state that although due to Variant B of the Weak Emergence Proposal one would hope to get a non-trivial correction, we are still meeting the requirements of
  Variant A and no direct contradiction to either of them is observed.

\subsection{Emergence in the asymptotic regime \texorpdfstring{$t_1\gg 1$}{TEXT}}
\label{sec_emerge_t}

Before we  discuss potential consequences of our result, we
consider other asymptotic regions in the moduli space.
In this section we focus on  the $t_1\to\infty$  limit.

\subsubsection*{One loop corrections to the moduli space metric}
  
It is a difficult question to decide  what kinds of bound states
exist for these wrapped branes and how the  final mass formula
for them reads.
In appendix \ref{app_c}, essentially by analogy we propose
an analogous mass formula as for the
perturbative string states \eqref{KKwindstrmassa}.
There, we also include the axions and realize the appearance
of the axion $\rho$ in the mass formula.
Using the relations from appendix \ref{app_c}, the computation for the
species scale and the one-loop kinetic terms proceeds
as before and we arrive at the analogous result

\eq{
            &G^{(1)}_{t_1 t_1}={\lambda}\,
            G^{(0)}_{t_1 t_1}\,,\qquad  G^{(1)}_{b_1
              b_1}=0\,, \qquad  G^{(1)}_{t_1
              b_1}=0\,, \\
       &G^{(1)}_{t_1 {\cal
                M}_a}=0\,, \qquad\quad\ \  G^{(1)}_{b_1 {\cal
                M}_a}=0\,,\\[0.1cm]
            \qquad
             &G^{(1)}_{{\cal M}_a {\cal M}_b}={2\lambda\over
              \log^2\!\big({M_{\rm pl}\over M_s}\big)}\, G^{(0)}_{{\cal M}_a {\cal M}_b}
          \,,
        }
with $M_{\rm pl}/M_s=\sqrt{t_1}$ and
 $a,b$ labelling all modes except $t_1$ and $b_1$.\,\footnote{Note that the
    KK-modes, the NS5-branes and  the last four
  $D2$-branes wrap torsion cycles of the orbifold, which
  would  be absent for a CY with $\pi_1(X)=0$. As a consequence, one
  would not get any non-trivial one-loop metric for the
  hypermultiplets in the latter case.}

\subsubsection*{One loop corrections to the gauge kinetic terms}

In contrast to the weak string coupling limit, now some of the light
states are wrapped $D$-branes and we can get a non-trivial one-loop
contribution to the gauge kinetic terms. 
As in the previous paragraph, we assume that the suggestive
mass formulas from appendix \ref{app_c} are correct so
that the computation can proceed similarly to the one from section \ref{sec_emergeKKWS}.
Let us only mention some of the main points here.

To evaluate the one-loop correction  \eqref{emergentgaugecoupling}, we first need
to identify those light states that are
charged under the four gauge symmetries.
Looking at table \ref{table_Dbranes} we realize that the $D0$-brane is electrically charged
under ${\cal A}^0$ and the first two wrapped $D2$-branes from that table
are electrically charged under ${\cal A}^2$ and ${\cal A}^3$, respectively.
There is no wrapped $D2$-brane that is electrically charged under
${\cal A}^1$, but there is the wrapped $D4$-brane, which is
magnetically charged under ${\cal A}^1$. This means that this brane
is electrically charged under the magnetic dual gauge field $\tilde
{\cal A}^1$. For the following, we have to keep in mind that the perturbative
one-loop correction \eqref{emergentgaugecoupling}  to the gauge coupling has been derived for
electrically charged particles running in the loop and we will
only apply it to such cases.

The relevant piece from the mass formula for these four types of
branes is precisely \eqref{masscontra} which we repeat here for convenience
\eq{
  \label{masscontrahere}
      M^2={M^2_{\rm pl}\over t_1} \Bigg[&\bigg(\overbrace{ {n_1
            +b_3 n_2 + b_2 n_3 + b_2 b_3 n_4\over  t_2^{1\over
              2} \,t_3^{1\over 2}  }}^{w_1}\bigg)^2+
        \bigg(\overbrace{ { (n_2 + b_2 n_4) \,   t_3^{1\over
              2} \over t_2^{1\over 2}  }}^{x_1}\bigg)^2+\\[0.2cm]
        &\bigg(\underbrace{ { (n_3 + b_3 n_4) \,   t_2^{1\over
              2} \over t_3^{1\over 2}  }}_{y_1}\bigg)^2+
         \bigg(\underbrace{ { n_4    t_2^{1\over
              2} \, t_3^{1\over 2}  }}_{z_1}\bigg)^2+\ldots\Bigg]\,.
  }
Say one wants to compute  $f^{(1)}_{00}$. 
As already observed in \cite{Castellano:2022bvr}, in the presence of a
non-trivial Kalb-Ramond field one has to recall   (see e.g. \cite{Grimm:2004ua})  that the 4D gauge
fields ${\cal A}^\Lambda$ are defined via the exact pieces in  $\hat F_2=dC_1$ and $\hat F_4=dC_3 - H_3\wedge
C_1$.  For the latter one expands $C_3=\sum_{I=1}^3 {\cal A}^I
 \wedge\omega_I$ so that  
the charges do not change and are still integer valued.
Hence, we have $q_0=n_1$ and the starting point of the computation is
\eqref{emergentgaugecoupling}
\eq{
  f^{(1)}_{00} \simeq 
\underbrace{\sum_{\vec{m}^I,\,\vec{n}^I,\,\vec{p}^I,\,N}}_{M\le \tilde\Lambda}
\!\!  {\rm deg}_N \, 
          (n_1)^2 \,,
 }
where the sum is over all states. As explained, we have set
all one-loop beta-coefficients to one. Let us stress that
the independence of the one-loop corrections from the coefficients $\beta,
\gamma$ and $\nu$ implies that even if  higher spin fields were
inducing an extra  polynomial or even exponential
$\exp(\beta'\sqrt{N})$ factor, they would remain unchanged. 

To avoid couplings to the
wrapped $D4$-branes, we set $b_2 b_3=0$ but allow
either of them to be non-zero.
After approximating  the  sum by
an  integral and expressing the ${\cal A}^0$ charge $n_1$ as
\eq{
               n_1=t_2^{1\over 2} t_3^{1\over 2}\left(  w_1 -
                 {b_3\over t_3}  x_1 - {b_2\over t_2} y_1 \right)\,,
}
one can proceed as in section \ref{sec_emergeKKWS}. Then one realizes that via
the integral over the 11 angular directions
$\varphi_1,\ldots,\varphi_{11\,}$, all off-diagonal
contributions from $n_1^2$ vanish so that only the diagonal ones
survive. Following the same steps as in the previous metric calculation, we finally arrive at
\eq{
  f^{(1)}_{00}\simeq
  t_1 t_2 t_3 \left( 1+
              \left({\textstyle {\dfrac{b_2}{
                  t_2}}}\right)^2 +  \left({\textstyle {\dfrac{b_3}{ t_3}}}\right)^2
        \right)
 {1\over 2\log\big({M_{\rm pl}\over M_s}\big)}   \,.
 }                
 This is proportional to the classical gauge coupling from \eqref{gaugeterms},
 with only the term involving $b_1$ missing. But that is expected,
 as $b_1$ does not appear in the mass formula for the light states.
 
 We can proceed similarly for the other electric components of the gauge
 coupling, i.e. $\Lambda,\Sigma=0,2,3$. For $\Lambda=\Sigma=1$ one can only determine the
 magnetically  dual coupling $\tilde f^{(1)}_{11}$ in terms of  $\tilde f^{(0)}_{11}=(f^{(0)}_{11})^{-1}$ .
 Now, the
 wrapped $D4$ branes (for $b_2=b_3=0$) are  electrically
 charged and we can still apply the formalism.
 In this manner,  the final result for the non-vanishing one-loop corrections to the gauge
kinetic functions can be compactly written as
\eq{
  \label{gaugecoupl1loop}
  f^{(1)}_{\Lambda\Sigma}&= {\xi\over
    2\log\big({M_{\rm pl}\over M_s}\big) }
            f^{(0)}_{\Lambda\Sigma}\Big\vert_{b_1=0} \ \
            (\Lambda,\Sigma=0,2,3)\,,\\
            \tilde f^{(1)}_{11}&=
            {\xi\over 2\log\big({M_{\rm pl}\over M_s}\big)} \tilde f^{(0)}_{11}
            \,, }
 where $\xi$ is a common numerical prefactor.

By electric-magnetic duality one might also extract some information
on the one-loop corrections to the theta-angles, but we stop here
and state that in the large K\"ahler modulus limit $t_1\to\infty$,
the loop corrections to the gauge couplings are essentially 
consistent
with Variant B of the Weak Emergence Proposal, but again miss some
dependence  (on $b_1$) and receive a suppression by
$\log(M_{\rm pl}/M_s)\sim \log(t_1)$.

\subsection{Emergence in the asymptotic regime \texorpdfstring{$u_1\gg 1$}{TEXT}}

For completeness, let us now discuss  emergence in the
limit $u_1\to\infty$.
If we again assume that we get an analogous mass formula as for the
perturbative string states \eqref{KKwindstrmassa}, then the computation for the
species scale and the one-loop kinetic terms will proceed
as before and we will arrive at the analogous result\,\footnote{Note that the
    two perturbative as well as all the
    wrapped $D2$, $D4$ and NS-branes wrap torsion cycles of the orbifold, which
    would  be absent for a CY with $\pi_1(X)=0$. According to \cite{Marchesano:2019ifh},
    in this case the classical infinite distance point in the complex structure
    moduli space could be obstructed by moving it  to finite distance by the presence of Euclidean
    $D2$-brane instantons. As described below table
      \ref{table_DbraneUs}, for this toroidal setting the large complex structure
      limit is related to the unobstructed large  K\"ahler modulus limit by two
      T-dualities. This suggests that the  $D2$-brane instantons do
      not carry the right fermionic zero mode structure to contribute
      to the hypermultiplet metric.}
      

  \eq{
            &G^{(1)}_{u_1 u_1}={\lambda}\,
            G^{(0)}_{u_1 u_1}\,,\qquad  G^{(1)}_{v_1
              v_1}=0\,, \qquad  G^{(1)}_{u_1
              v_1}=0\,, \\
       &G^{(1)}_{u_1 {\cal
                M}_a}=0\,, \qquad\qquad\, G^{(1)}_{v_1 {\cal
                M}_a}=0\,,\\[0.1cm]
            \qquad
             &G^{(1)}_{{\cal M}_a {\cal M}_b}={2\lambda\over
              \log^2\!\big({M_{\rm pl}\over M_s}\big)}\,G^{(0)}_{{\cal M}_a {\cal M}_a}
          \,,
        }
with $M_{\rm pl}/M_s=\sqrt{u_1}$
and  $a,b$ labelling all modes except $u_1$ and $v_1$. Since all states in the light towers are neutral with respect to the
graviphoton and the other three $U(1)$ gauge symmetries, the one-loop
corrections to their gauge couplings do vanish.

\subsection{Consequences for Emergence}

In view of the discrepancy of the BH and the QFT
approach in determining the species scale for string towers,
let us now discuss   what the consequences of these results 
for the emergence proposal could be.
All the one-loop
corrections  have been derived in the QFT picture and
as such in particular the $\log$-factors are under scrutiny.
There is the  possibility that the
extrapolation of the QFT techniques to the stringy regime
with energies larger than $M_s$ is at least questionable.

However, in the course of the computation, we have seen
a certain universality in  the final results for the
emerging field metric and gauge couplings, which first of all  
show the expected classical moduli dependence in some
  of the orthogonal  components.
The latter arose  from the inclusion of both the KK and
the winding modes, which in the regime of interest
are already heavier than the string scale.
In addition, we observed this  highly  non-trivial conspiracy
of  coefficients in the relation \eqref{wonderrel},
which we think is not just coincidental.

Thus,  in the remainder of this section we approach our results
with a positive attitude and discuss two possibilities
to interpret the appearing $\log$-factors.

\subsubsection*{A: The log-factors are unphysical}

Motivated by the BH picture we just ignore the $\log$-suppressions.
Going through the computation, one realizes  that they really entered
the expression for the one-loop field metric and gauge coupling
by inserting $\tilde\Lambda/M_s\sim \log(\sigma)$ in the final step
(see e.g. the discussion around equation \eqref{UTmetric}).

In the weak string coupling limit, we would say that the field
metric is almost fully emerging with only the $G_{\rho\rho}$ term
missing. This is due to the non-appearance of the axion $\rho$
in the mass formula for the light towers of states and seems to be
at odds with supersymmetry.
For the non-vanishing metric components, the relative normalizations are also fine, except
for a factor of 2 between  $G_{\sigma\sigma}$ and the other
components.

Since the one-loop gauge couplings were all trivially
vanishing, clearly these do not emerge. As mentioned,
on a genuine CY with vanishing fundamental group, also
the 1-loop metric components for the K\"ahler moduli would be vanishing.
Essentially, this is just a reflection of the non-mixing
of hyper and vector multiplets in 4D $N=2$ supergravity.


Consistent with $N=2$ supersymmetry, 
in the limit of a single large K\"ahler modulus, also some components
of the gauge couplings emerged with the expected moduli dependence.

\subsubsection*{B: The log-factors are physical}

The second possibility is that the $\log$-factors are physical
and should be taken seriously.
For concreteness, let us only consider the example
of the weak string coupling case.
As in the weak version of the Emergence Proposal,  including the
classical contributions,  we have up to the one-loop level
\eq{
  \label{kinetot}
        {\cal L}_{\rm kin}=&M_{\rm pl}^2\left({1\over \sigma^2}+{\lambda\over
            \sigma^2 }+\ldots\right) (\partial\sigma)^2+ {M_{\rm pl}^2\over 4\sigma^4}(\partial\rho)^2\\[0.1cm]
   &\phantom{aaaaa}+ M_{\rm pl}^2\sum_I\left({1\over 4t_I^2}+{\lambda\over 2t_I^2 \log^{2}(\sigma)}+\ldots\right) \Big((\partial
     t_I)^2+(\partial b_I)^2\Big)
 } 
 and a similar term for the complex structure moduli ${\cal U}_I=v_I+iu_I$.
 The dots indicate that there will be higher order corrections, as in
 our computation we were just extracting the asymptotic form of
 the loop corrections.

\begin{figure}[ht]
    \centering
    \includegraphics[width=0.8\textwidth]{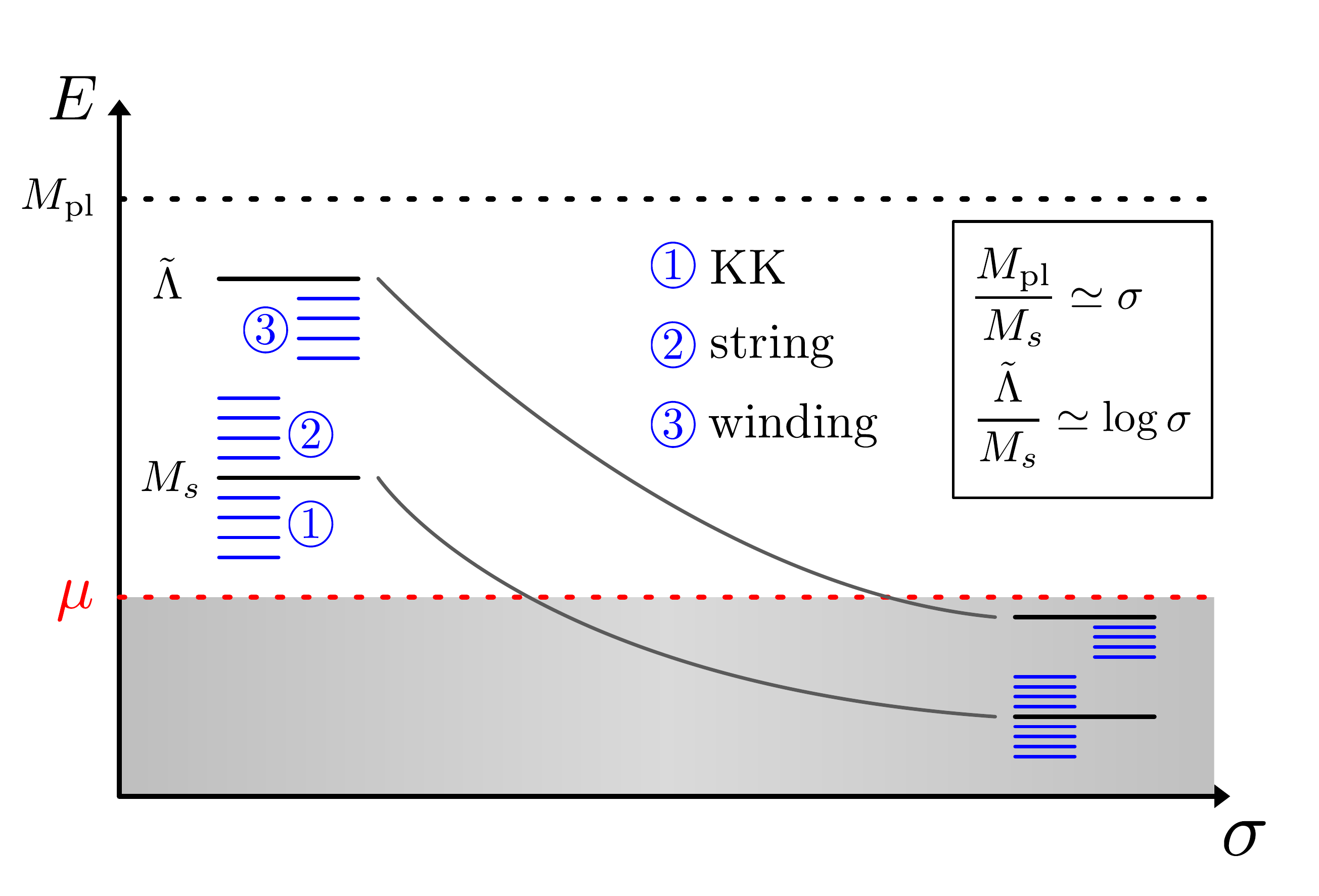}
    \caption{Dependence of energy scales 
      on $\sigma$. The Swampland Distance Conjecture predicts a
      breakdown of the EFT at large (Planck scale) field distances when many states
      fall below its cut-off $\mu$.
    }
    \label{scales_sigma}
\end{figure}

We first notice that \eqref{kinetot} indicates that the 
induced corrections due to the light towers of states appearing in
asymptotic limits in field space are much milder than initially
advocated. Against the intuition from the Swampland   Distance Conjecture that the
EFT becomes worse and worse with more and more states dropping below
the UV cut-off (see also figure \ref{scales_sigma}), in the
infinite field distance limit the one-loop corrections are either
proportional  to the classical values or
are subleading.
However, this does not happen polynomially in $\sigma^{-1}=\exp(-\phi_4)$
like in perturbation theory, but logarithmically with a suppression
$\log^{-2}(\sigma)$. Thus, in this sense
the (kinetic) couplings of the classical EFT, in which
one is only keeping the generically lightest states, are still valid
and under control.

However, the correction is moduli dependent and one might wonder
about the initial $N=2$ supersymmetry.
As we have already mentioned, in a 4D EFT with $N=2$ supersymmetry,
the moduli spaces of the vector- and the hyper multiplets are not
allowed to mix. But since the $\sigma$-field resides in the type IIA
hyper multiplet and $\sigma$-dependent corrections appear in the
one-loop metric of the Kähler moduli (belonging to the vector
multiplet), it seems that the two sectors do not decouple in our
case. 

A natural way to resolve this issue is to interpret the problematic
$\log$-factor in $G^{(1)}_{t_I t_I}= M^2_{\rm pl}\, g^{(1)}_{t_I t_I}$ not
as a correction to the field metric, but as a one-loop correction to
$M_{\rm pl}^2$. Recall that the latter is nothing else than the
kinetic term for the graviton, which should/could also emerge. 
Indeed, the stringy one-loop correction for the type II string on
4D orbifold models with $N=2$ supersymmetry was found
to be non-vanishing in \cite{Kiritsis:1994ta,Kohlprath:2002fe}
and in fact proportional to the string scale, i.e.
$\delta( M_{\rm pl}^2)\sim M_s^2$.
With this modification, the expansions of the relevant quantities read
\eq{
   M_{\rm pl}^2&=M^{2\,(0)}_{\rm pl} + M^{2\,(1)}_{\rm pl}+\ldots\\
   g_{AB}&=   g_{AB}^{(0)}+ g_{AB}^{(1)}+\ldots \;,
}
where $g_{AB} = G_{AB}/M_{\rm pl}^2$. The previously computed one-loop coefficients therefore consist of two different components, namely
\eq{
      \label{one-loop-general}
      G_{AB}^{(1)}=M^{2\,(0)}_{\rm pl}\,  g_{AB}^{(1)}+ M^{2\,(1)}_{\rm pl}\,g_{AB}^{(0)} \,.
}
Let us now apply \eqref{one-loop-general} to the $\log$-corrected metric components of the Kähler moduli in \eqref{metric_full_string}. Here, the $\sigma$-dependent factor corrects the Planck mass according to
\eq{
              M^{2\,(1)}_{\rm pl}={2 \lambda\, M^{2\,(0)}_{\rm
                  pl}\over \log^{2}(\sigma) } \simeq
                    N_{\rm sp} \, M_s^2 \,,
                  }
which for a small number of light species is consistent with \cite{Kiritsis:1994ta,Kohlprath:2002fe}.
Now the one-loop corrections to the field metrics of the
orthogonal directions  are all vanishing
\eq{
  g^{(1)}_{ab}=0\,,\qquad  {\rm with}\quad  a,b=t_I,b_I,u_I,v_I\,.
}
As a consequence, the found emergence-like
relation $G^{(1)}_{ab} \sim G^{(0)}_{ab}/\log^2(\sigma)$ is just a trivial
  consequence of \eqref{one-loop-general}.
Next, determining the corrections to the remaining metric components
we find
\eq{
  \label{hannover96}
         g_{\sigma\sigma}^{(1)}&={\lambda \over \sigma^2} - {2 \lambda
           \over \log^{2}(\sigma)\sigma^2} \,,\qquad\quad
         g_{\rho\rho}^{(1)}=-{\lambda \over 2\log^{2}(\sigma)\sigma^4} \,,
}
so that  only the hyper multiplet metric receives logarithmic
hyper multiplet depending corrections. 
However, the aforementioned asymmetry with respect to the saxion $\sigma$ and its
axionic partner $\rho$ persists.

 
\section{Conclusions}

In this paper we were exploring  the Emergence Proposal with  a
concrete $N=2$ supersymmetric toroidal orbifold model where, in particular in the weak string coupling limit, we had full
control over the light particle and string towers of states and  their detailed mass formula.
We were able to carry out  the computation 
keeping track of all 14 moduli fields from the NS sector.
We also considered two other
asymptotic regimes, namely  the  large K\"ahler and large complex
structure limits, where the same number and mass pattern appeared
once one included all light modes. Indeed, we identified
12 light particles mostly arising from wrapped $D$-branes and wrapped
NS-branes and one low-tension 4D string arising from a wrapped
NS-brane.

Concerning the string tower, we pointed out an issue
with the definition of the species scale, namely the QFT and the
BH picture were giving mutually non-consistent results.
From the QFT point of view, this might be rooted in the fact
that with an emergent string present in the asymptotic field limit,
one inevitably probes the string at energies larger than $M_s$.
The most important question clearly is to completely resolve this issue
and in the course develop a well-founded approach to treat the Emergence 
Proposal in the presence of asymptotically tensionless strings.

As the only available approach, carrying out the  self-consistent computation in
the QFT picture, we
found that the one-loop corrections to the
moduli field metric and the gauge couplings follow
a very similar pattern as their classical results.
However, the details turned out to be more intricate
and we discussed essentially two different ways
to interpret the results differing in how we 
treated the ubiquitously appearing $\log$-suppressions. 
Just ignoring them, a large portion of the classical
field metric and gauge couplings emerged, i.e.
they in particular had the expected moduli dependence.
No components were generated at one-loop that
were absent classically.

The other option was to consider the $\log$-factors
as being physical.
Avoiding any conflict
with the decoupling of hyper- and vector multiplets
in the $N=2$ supersymmetric EFT led us to the
inclusion of a moduli dependent one-loop
correction to the 4D Planck mass.
However,  taken at face value, we have seen 
that neither of the two variants of the Weak Emergence Proposal
is really fully satisfied.

As another new aspect of emergence, this interpretation suggested
that the potentially induced corrections due to the light towers
of states appearing in asymptotic limits in field space are much
milder than initially advocated.

To get further confirmation, it would  be interesting to further generalize our
multiple moduli computation.
Of course, one could consider many other infinite
distance limits in the saxionic moduli space, like e.g. the overall
large
volume limit, where one scales all K\"ahler moduli like $t_I=\lambda
\hat t_I$ with the 4D dilaton kept constant.
This is nothing else than the decompactification limit
of the dual M-theory on $CY\times S^1$ compactification. 
Such a  large volume  limit has been
one of  the prime examples  of emergence discussed
in\cite{Grimm:2018ohb,Castellano:2022bvr}, where the lightest states
were the $D0$-branes with the species scale being the 5D Planck-scale
$\tilde\Lambda \sim M_{pl}/ \lambda^{1\over 2}$.
In addition, one could consider 
backgrounds in other space-time dimensions or with less supersymmetry.

To resolve this issue about the correct value of the
species scale one might wonder whether one could carry out the full
emergence computation directly in string theory, i.e. without
employing the field theory diagrams.  For instance, in an asymptotic
direction in the $N=2$ vector multiplet moduli space  one
starts with  the proposal \cite{vandeHeisteeg:2022btw} that the species scale is related
to the topological one-loop free energy as $\tilde\Lambda\sim M_{\rm pl}/\sqrt{F_1}$ and
then involves also  the one-loop gauge thresholds corrections.

\vspace{0.2cm}

\noindent
\paragraph{Acknowledgments:}
We would like to thank Niccol\`o Cribiori, Christian Knei\ss l and Andriana Makridou for
discussions
and Max Wiesner for sharing his insights about the species scale of
string towers with us.
The work of R.B. and A.G. is funded by the Deutsche Forschungsgemeinschaft (DFG, German Research Foundation) under Germany’s Excellence Strategy – EXC-2094 – 390783311.

\vspace{0.4cm}
\clearpage
\appendix

\section{Generalities on emergence}
\label{app_a}

To be self-contained, we collect here the basics of the Emergence
Proposal, providing in particular  the background material and
formulas that are needed in the body of the paper.

As already mentioned, in the weaker version of the Emergence Proposal
it is claimed that any infinite  field distance singularity can be
associated to a tower of light states becoming massless in that
limit. The tree level singular behavior of the metric can then be
matched by integrating out these states up to the  quantum gravity
cut-off, namely the  species scale $\tilde{\Lambda}$.
In general, we have
multiple towers of states with moduli dependent mass scales  and a further
degeneracy in the mass spectrum. 
Therefore, the number of states $N_{\rm sp}$ is given by
\begin{equation}
  \label{Nsp general}
    N_{\rm sp}=\sum_{\vec{n}}{\rm deg}_{\vec{n}}\,,
\end{equation}
where we are summing over all quantum numbers $n_i$, collectively denoted as $\vec{n}$, such that
the corresponding states have masses below the species scale. Note
that \eqref{Nsp general} is valid in the case of multiplicative
towers, which we focus on, where states of mixed quantum numbers
are allowed to appear. If that is not the case, then we have what is
called additive towers, where $N_{\rm sp}=\sum_{i}N_{\rm sp,i}$ \cite{Castellano:2022bvr}.
If the  species scale is much bigger than the mass scales
of all towers, one can employ an  integral approximation
and can often carry out these higher dimensional integrals
either analytically or at least extract their  asymptotic form.

For emergence, we are for instance interested in the kinetic terms for the moduli
fields, whose classical action takes the form
\eq{
  \label{classical metric}
    S_{\rm kin} = -\frac{1}{2} \int d^4 x\sqrt{-g} \, \underbrace{G^{(0)}_{ab} \,\partial_\mu
    \phi^a \,\partial^{\mu} \phi^b}_{{\cal L}_{\rm kin}}\,.
  }
The above conventions correspond to dimensionless fields with
$c=\hbar=1$ and in 4D Minkowski space we have of course
$g_{\mu\nu}=\eta_{\mu\nu}$.
Note that in our conventions the classical field metric $G^{(0)}_{ab}$
contains a factor of $M_{\rm pl}^2$.
Qualitatively speaking, emergence means that the one-loop contribution  $G^{(1)}_{ab}$ to this moduli
field metric arising from integrating out the aforementioned  light species is
proportional to the tree-level metric.
Closely following
\cite{Castellano:2022bvr,Grimm:2018ohb}, consider a tower of massive
real scalars $\varphi_{\vec{n}}$ or Dirac fermions $\psi_{\vec{n}}$, whose
mass is parametrized by the  moduli fields $\phi^a$ and some quantum
numbers $n_i$. 
This is captured by
the mass term of the Lagrangian
\eq{
  \label{mass lagrangian}
    \mathcal{L}^B_{\rm mass} = M_{\rm pl}^2\sum_{\vec{n}} \frac{1}{2} m_{\vec{n}}^2(\phi^a) \,\varphi_{\vec{n}}^2\qquad\text{or}\qquad \mathcal{L}^F_{\rm mass}=M_{\rm pl}^3\sum_{\vec{n}}m_{\vec{n}}(\phi^a) \,\overline{\psi}_{\vec{n}} \,\psi_{\vec{n}} \,.
  }

\begin{figure}[ht]
    \centering
    \includegraphics[width=\textwidth]{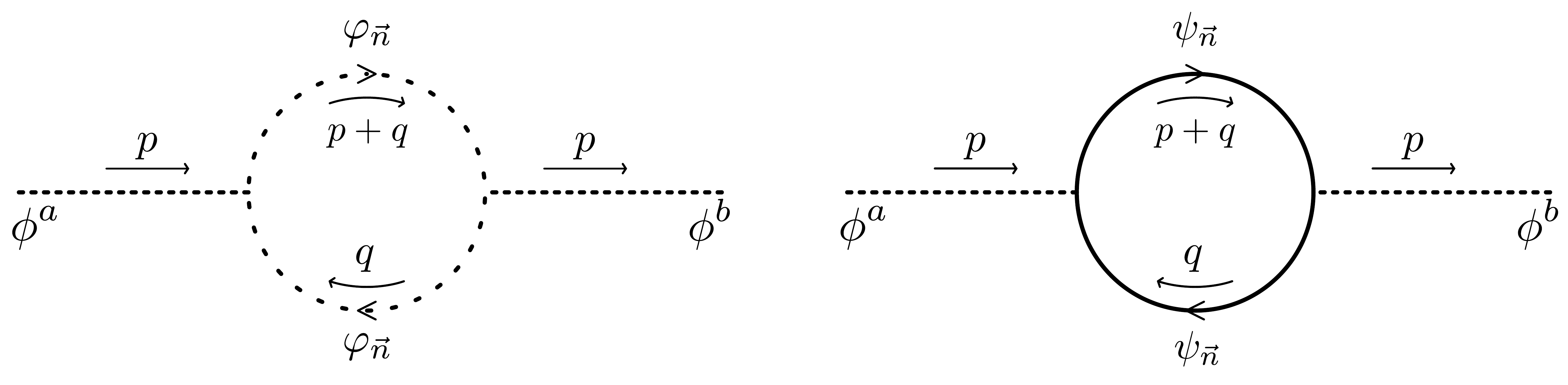}
    \caption{Feynman diagrams for the processes leading to the one-loop corrections to the field metric, with massive scalars $\varphi_{\Vec{n}}$ (left) or massive fermions $\psi_{\Vec{n}}$ (right) running in the loop. 
    }
    \label{one-loop-moduli}
\end{figure}

The expansion of each mass term to linear order in the perturbations
around the vacuum expectation values  of the fields $\phi^a$ results in a trilinear interaction vertex with coupling strengths
\eq{
    \label{coupling}
    \lambda_{a,\vec{n}} = 2 m_{\vec{n}} \del_{a} m_{\vec{n}} \; {(\rm scalars)}  \qquad {\rm or} \qquad  \mu_{a,\vec{n}} = \del_{a} m_{\vec{n}} \; {(\rm fermions)} \,
  }
 with  $\partial_a=\partial/\partial\phi^a$.
These vertices lead to the Feynman diagrams shown in figure \ref{one-loop-moduli}.

Upon integrating out the states from the tower, the propagator matrix $D_{ab}(p^2)$ of the moduli receives a one-loop correction given by
\eq{
    D_{ab}(p^2) = \frac{1}{p^2 - \Pi_{ab}(p^2)} \,, \qquad \Pi_{ab}(p^2) = \sum_{\vec{n}} \, \Pi_{ab,\vec{n}}(p^2) \,.
}
 Here,  $\Pi_{ab,\vec{n}}(p^2)$ is the contribution of a single amputated one-loop Feynman diagram containing the boson or fermion of the tower characterized by $\vec{n}$ and the index structure is due to \eqref{coupling}. The resulting wave-function renormalization of the moduli is equivalent to the one-loop metric we are looking for. Since it is given by the part of $\Pi_{ab}(p^2)$ proportional to $p^2$, we simply need to take the derivative of each $\Pi_{ab,\vec{n}}(p^2)$ with respect to $p^2$, evaluate at $p=0$ and sum over the whole spectrum. The one-loop metric comes out as
\eq{
    \label{one-loop}
    G_{ab}^{(1)} =
    \sum_{\vec{n}} \frac{\del \Pi_{ab,\vec{n}}(p^2)}{\del p^2} \Big\rvert_{p^2 = 0} \,.
}
With the conventions of figure \ref{one-loop-moduli}, the contribution from a scalar loop reads
\eq{
    \Pi_{ab,\vec{n}}(p^2) = \frac{\lambda_{a,\vec{n}}\,\lambda_{b,\vec{n}}}{2} \int \frac{d^4 q}{(2\pi)^4} \frac{1}{q^2 + m_{\vec{n}}^2} \frac{1}{(p+q)^2 + m_{\vec{n}}^2} \,,
}
where the factor of $1/2$ accounts for the symmetry of the bosonic diagram. For the contribution to the one-loop metric one obtains
\eq{
    \label{wf_renorm_boson_1}
    \frac{\del \Pi_{ab,\vec{n}}(p^2)}{\del p^2} \Big\rvert_{p^2 = 0}
    &= - \frac{\lambda_{a,\vec{n}}\,\lambda_{b,\vec{n}}}{2} \int
    \frac{d^4 q}{(2\pi)^4} \frac{1}{(q^2 + m_{\vec{n}}^2)^3} \\
    &=- \frac{\lambda_{a,\vec{n}}\,\lambda_{b,\vec{n}}}{16\, (2\pi)^2}
    {\tilde\Lambda^4\over
        m_{\vec{n}}^2 ( \tilde\Lambda^2+m_{\vec{n}}^2)^2}\,.
}
The momentum integral was performed up to the species scale $\tilde
\Lambda$, since only those light modes will be included in the
emergence calculation.
Now, one can distinguish two asymptotic limits:
either $\tilde \Lambda \gg m_{\vec{n}}$, which is typically fulfilled
by KK towers, or $\tilde \Lambda \simeq m_{\vec{n}}$, which holds for
most states in a tower of string excitations. Apparently, both limits
give the same functional behavior, namely
\eq{
    \label{wf_renorm_boson_2}
    \frac{\del \Pi_{ab,\vec{n}}(p^2)}{\del p^2} \Big\rvert_{p^2 = 0} \simeq \frac{\lambda_{a,\vec{n}}\,\lambda_{b,\vec{n}}}{m_{\vec{n}}^2}
  }
and only the overall numerical coefficient is different. As
 mentioned, our computation is indifferent to such overall factors
 so that the  form \eqref{wf_renorm_boson_2} is sufficient for our purposes.

Fermionic loop integrals can be computed in a similar way. The Feynman
diagram on the right hand side of  figure \ref{one-loop-moduli} gives
\eq{
    \Pi_{ab,\vec{n}}(p^2) = -\mu_{a,\vec{n}}\,\mu_{b,\vec{n}} \int \frac{d^4 q}{(2\pi)^4} {\rm tr} \left( \frac{(-i \slashed{q}+ m_{\vec{n}})(-i(\slashed{p} + \slashed{q}) + m_{\vec{n}})}{(q^2 + m_{\vec{n}}^2)((p+q)^2 + m_{\vec{n}}^2)}\right) \,.
}
With the trace in the above integral explicitly performed, the part
linear in $p^2$ splits into the  two pieces
\begin{eqnarray}
    \label{wf_renorm_fermion_1}
    \frac{\del \Pi_{ab,\vec{n}}(p^2)}{\del p^2}\Big\rvert_{p^2 = 0}=&-\underbrace{4 \,\mu_{a,\vec{n}}\,\mu_{b,\vec{n}}\, \int\frac{d^4 q}{(2\pi)^4} \frac{1}{(q^2+m_{\vec{n}}^2)^2}}_{\equiv \, (I)}\\[0.1cm]
    &+\underbrace{2 \cdot 4 \,\mu_{a,\vec{n}}\,\mu_{b,\vec{n}} \, \int\frac{d^4 q}{(2\pi)^4} \frac{m_{\vec{n}}^2}{(q^2+m_{\vec{n}}^2)^3}}_{\equiv \, (II)}\,,
\end{eqnarray}
where the factor 4 comes from the trace of gamma matrices and counts
the number of fermionic degrees of freedom. For $(II)$, one finds the
exact same behavior of the loop-integral as in the bosonic case, only
with opposite sign. We focus on supersymmetric setups, where the
number of on-shell bosonic and fermionic degrees of freedom match, so
these terms precisely cancel out.
Carrying out the $q$-integration, 
  for  the remaining contribution $(I)$ in \eqref{wf_renorm_fermion_1} one obtains
\eq{
    \label{wf_renorm_fermion_inti}
    \frac{\del \Pi_{ab,\vec{n}}(p^2)}{\del p^2}\Big\rvert_{p^2 = 0} =
    - \,{\mu_{a,\vec{n}}\,\mu_{b,\vec{n}}\over (2\pi)^2} \left(
        \log\bigg( {\tilde\Lambda^2+ m_{\vec{n}}^2\over m_{\vec{n}}^2}\bigg)
        -{\tilde\Lambda^2\over \tilde\Lambda^2 +m_{\vec{n}}^2}\right)\,.
   } 
In the KK-like limit $\tilde \Lambda \gg m_{\vec{n}}$, this
becomes 
\eq{
    \label{wf_renorm_fermion_2}
    \frac{\del \Pi_{ab,\vec{n}}(p^2)}{\del p^2}\Big\rvert_{p^2 = 0} \simeq -\mu_{a, \Vec{n}}\, \mu_{b, \Vec{n}}  \,\log \left( \frac{\tilde{\Lambda}^2}{m_{\Vec{n}}^2}\right)\,.
  }
The one-loop metric is given by the sum of these contributions. Inserting \eqref{coupling} and \eqref{wf_renorm_fermion_2} in \eqref{one-loop} and taking into account possible mass degeneracies, we arrive at
\eq{
    \label{one-loop-KK}
    G_{ab}^{(1)} \simeq
    \sum_{\vec{n}} {\rm deg}_{\vec{n}} \, \del_{a} m_{\vec{n}} \, \del_{b} m_{\vec{n}} \, \log \left( \frac{\tilde{\Lambda}}{m_{\Vec{n}}}\right)\,.
}
For the string-like limit $\tilde \Lambda \simeq m_{\vec{n}}$, the
contribution   \eqref{wf_renorm_fermion_inti} reduces to the already familiar expression \eqref{wf_renorm_boson_2} and leads us to the one-loop metric
\eq{
    \label{one-loop-string}
    G_{ab}^{(1)} \simeq
     \sum_{\vec{n}} {\rm deg}_{\vec{n}} \; \del_{a} m_{\vec{n}} \, \del_{b} m_{\vec{n}} \,.
}
Strictly speaking, these relations are derived for scalars (spin-0)  and
spin-1/2 fermi\-ons. Since we can also include towers of string
excitations, there can be contributions from higher spin
bosons and fermions, as well. We assume
that their contribution per degree of freedom will not essentially
deviate from the lowest spin cases discussed. Hence,  we
apply the one-loop metrics \eqref{one-loop-KK} and \eqref{one-loop-string} also to these higher
spin states.\footnote{After all, the contribution from the string excitations was very robust and independent of  a couple of parameters. Therefore, one can probably even weaken this assumption.}

\subsubsection*{Gauge kinetic terms at one-loop}

\begin{figure}[ht]
    \centering
    \includegraphics[width=\textwidth]{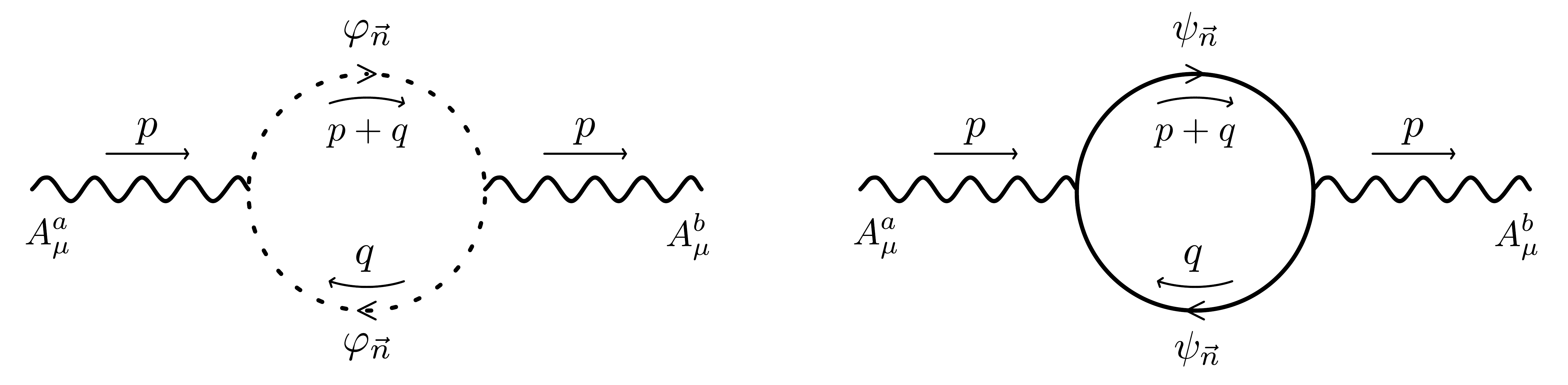}
    \caption{Feynman diagrams for the processes leading to the one-loop corrections to the gauge kinetic terms, with massive scalars $\varphi_{\Vec{n}}$ (left) or massive fermions $\psi_{\Vec{n}}$ (right) running in the loop.
    }
    \label{one-loop-gauge}
\end{figure}

The emergence idea can be similarly utilized to study the behavior of
gauge couplings in the infrared. Let us briefly sketch the logic for a
set of $U(1)$ gauge fields $A_{\mu}^a$ with field strengths $F_{\mu
  \nu}^a = 2 \del_{[\mu} A^a_{\nu]}$ in 4D, allowing for classical
gauge-kinetic mixings according to
\eq{
  \label{classicalgauge}
  S_{\rm kin} =
  \int d^4 x\sqrt{-g} \,
      \left(-{1\over 4}\sum_{a,b}  f^{(0)}_{ab} \, F^{\mu \nu,a} F_{\mu \nu}^b\right)
}
with  the gauge kinetic function given in terms of the
gauge couplings as  $f^{(0)}_{ab}=g^{-2}_{ab}\,$. Once again, we
incorporate towers of scalars $\varphi_{\Vec{n}}$ and fermions
$\psi_{\Vec{n}}$ with mass $m_{\Vec{n}}$. They are minimally coupled
to the gauge fields via the covariant derivative 
\eq{
    D_{\mu} \varphi_{\Vec{n}} = (\del_{\mu} - iq_{a,\Vec{n}} \,A_{\mu}^a)\, \varphi_{\Vec{n}} \,, \qquad D_{\mu} \psi_{\Vec{n}} = (\del_{\mu} - iq_{a,\Vec{n}} \,A_{\mu}^a)\, \psi_{\Vec{n}}
}
where $q_{a,\Vec{n}}$ are   the electric charges.   
The propagator of the $U(1)$ fields receives corrections from analogous loop diagrams depicted in figure \ref{one-loop-gauge}, so the corrected expression reads
\eq{
    D_{ab}^{\mu \nu} (p^2) = \left( \frac{p^2}{g_{ab}^2} \delta^{\mu \nu}_{ab} - \Pi^{\mu \nu}_{ab}(p^2) \right)^{-1} \,, \quad \Pi^{\mu \nu}_{ab}(p^2) = \Pi_{ab}(p^2) \delta^{\mu \nu} \,,
}
where we are assuming Lorentz gauge for all vector fields, $\del^{\mu} A^a_{\mu} = 0$, and a flat background with Euclidean metric $\bar{g}_{\mu \nu} = \delta_{\mu \nu}$. As before, the kinetic terms are given by wave-function renormalizations, so one has to sum over all contributions from amputated one-loop diagrams.

Similar to the kinetic terms for  the moduli, one part from the fermionic loop
integrals always cancels the scalar contribution if supersymmetry is
unbroken. Likewise, both two interesting limits $\tilde \Lambda \gg
m_{\vec{n}}$ and $\tilde \Lambda \simeq m_{\vec{n}}$ lead to the same
asymptotic behavior as before. In view of our application, we only focus on the latter case, for which 
the total one-loop gauge couplings are  given by
\eq{
  \label{emergentgaugecoupling}
   f^{(1)}_{ab}  \simeq \sum_{\vec{n}} \,{\rm deg}_{\vec{n}} \; q_{a,\Vec{n}} \, q_{b,\Vec{n}} \,.
}
Analogous to the moduli metric, we will assume
that the  contribution per degree of freedom from higher spin bosons
and fermions will not essentially deviate. Hence, we set all the usually
appearing one-loop beta-function coefficients to one.

\subsubsection*{Cancellation of $\hbar$-factors}

Following \cite{Castellano:2022bvr}, let us comment on an important conceptual point. Although our starting point was accounting for quantum corrections, our results do remain classical. This subtlety is partially concealed because in natural units $M_{\rm pl}^2=\kappa_4^{-2}$, where $\kappa_4$ is Einstein's gravitational coupling constant. Restoring $\hbar$ in the graviton
 self energy expression \eqref{graviton self energy}   we  have\footnote{We will still set $c=1$. Recall that we can express the d-dimensional Planck mass as  $M_{\rm
     pl,d}^{d-2}=\hbar^{d-3}c^{5-d}/8\pi G_N$, where $G_N$ is Newton's
   gravitational constant  and $G_N=\kappa_d^2c^4/8\pi$.} 
 \begin{equation}
  \label{hbarfirst}
      \pi^{-1}(p)=p^2\left(1-\frac{N_{\rm sp}\,p^2\kappa_4^2}{120\pi\hbar}\log\left(-\frac{p^2}{\mu^2}\right)+\ldots\right)\,,
 \end{equation}
where for convenience we left out the mass dependent corrections.
 Noticing that the classical factor $\kappa_4^2$ comes from the
 graviton vertex, we realize that this one-loop diagram comes with an extra factor $\hbar^{-1}$.
 Now to illustrate our point and take advantage of \eqref{Nsp
   general}, let us consider the example of a tower of states with
 mass $n\,\Delta m$ with polynomial degeneracy $n^K$, corresponding to
 $N_{\rm sp}=n_{\rm max}^{K+1}$, where $\tilde{\Lambda}=n_{\rm
   max}\Delta m$.
 Using \eqref{one-loop-string}, we get 
\eq{
    G_{ab}^{(1)} \simeq \frac{1}{\hbar}n_{\rm max}^{K+3}\,
    (\del_{a} \Delta m)\,(\del_{b} \Delta m)\,.
}    
As for the one-loop amplitude \eqref{hbarfirst},
we have  introduced  an extra loop-factor of
$\hbar^{-1}$. Then the correction to the metric   can be expressed as
\eq{
 G_{ab}^{(1)} \simeq \frac{1}{\hbar}N_{\rm
   sp}\,\tilde{\Lambda}^2\frac{\del_{a} \Delta m\,\del_{b} \Delta
   m}{\Delta m^2}\simeq \frac{M_{\rm pl}^2}{\hbar} \frac{\del_{a}
   \Delta m\,\del_{b}  \Delta m}{\Delta m^2}\,,
}
which due to the relation $ M_{\rm pl}^2/\hbar=\kappa_4^{-2}$ gives
a classical result, where $\hbar$ has canceled out.
This is possible, as the cut-off is also defined via a one-loop
diagram.

\section{Species scale for KK modes}
\label{app_b}

Assume for simplicity that we have a circle compactification leading to a non-degenerate one-dimensional KK tower with a mass spacing
\begin{equation}
    \Delta m={M_s\over r}=\frac{M_{\rm pl}}{\sigma r}\,,
\end{equation}
where $r$ is the radius of the circle in string length units and
$\sigma$ the inverse of the 4D string coupling. The mass of each KK mode will be given by
\begin{equation}
    m_n=n\,\Delta m
\end{equation}
so that  the heaviest KK mode in our theory will be for $k=\tilde\Lambda/\Delta m$.
In the QFT approach, the total number of states up to level $k$ will be equal to $N_{\rm sp}=k$ so that invoking
also the definition of the species scale we get
\eq{
    \tilde{\Lambda}=\frac{M_{\rm pl}}{(\sigma  r)^{\frac{1}{3}}}\,,
        \qquad\qquad
    N_{\rm sp}=(\sigma r)^{\frac{2}{3}}\,.
 }
Note that the species scale is nothing else than the 5D Planck-scale.
      
Now let us consider the same model using the BH approach\footnote{We
  are   indebted to Niccol\`o Cribiori for contributing in an essential
  way to this computation.}, where the
calculation is more involved. In particular, while
$k=\tilde\Lambda/\Delta m$ still holds, we now  need to
count the number of multiparticle states whose total mass is equal to
the one of the  BH. This will be described by a (large) number $N$, given by
\begin{equation}
    \label{BH_KK_1}
    N=\frac{M_{BH}}{\Delta m} = S_{BH} \,k= N_{\rm sp} \, k\,.
  \end{equation}
where we have used the formula for the Bekenstein-Hawking entropy.  
Hence, their total number will
be given by the number of possible partitions of $N$ into numbers
smaller than $k$. Since $k\ll N$, we can approximate this number by
\begin{equation}
    \Omega_{N,k}\sim \frac{N^{k-1}}{(k-1)!\,k!}\,.
  \end{equation}
One realizes that for $N_{\rm sp}\sim k$ this takes the
asymptotic form
\eq{
        \Omega_{N,k}\sim \left({k^k\over k!}\right)^{2}\sim e^{2k}
      }
 where we have used the leading exponential term in Stirling's formula
 $k!\sim (k/e)^k$.     
Taking now the logarithm of this expression we find
\eq{
    \label{BH_KK_2}
    S=\log \Omega_{N,k}\sim k \,.
  }
  In the BH picture this is supposed to be equal to $N_{\rm sp}$,
  which is indeed consistent with our assumption.
  In conclusion, for KK modes both the QFT and the BH picture give
  the same values for the species scale and the number of light species.


\section{Axions in  the large K\"ahler modulus limit}
\label{app_c}

Relating to the discussion in section \ref{sec_emerge_t}, in this appendix we
speculate about the full mass formula
in the $t_1\to\infty$ limit.
So far, we were not explicitly including the axions in the discussion.
To do so, let us first consider the wrapped $D$-brane states from
table \ref{table_Dbranes}. Up to the overall factor
$M_{\rm pl}/\sqrt{t_1}$, these states involve only the
K\"ahler moduli $t_2,t_3$ and the complex structure moduli $u_2,u_3$
on the second and the third $T^2$.  Starting with the $D4$-brane,
turning on $b_2$ (or $b_3$) via the Born-Infeld action one generates
also a contribution to the tension that scales precisely like
the first $D2$-branes from that table. Turning on both $b_2$ and $b_3$
one gets a contribution like the $D0$-brane.
Hence, these four branes are related via turning on  non-trivial axion
values $b_2$ and $b_3$.

Note that such chains of four states were
also present in the original perturbative mass formula \eqref{KKwindstrmassa}.
It is a yet not  resolved issue  which  bound states of all these
(relatively non-supersymmetric) wrapped branes can form and what their mass is, so we just
speculate that the final form will be very similar to the
one for the perturbative states. The shift symmetry of the axions
provides some constraints. Hence, it is suggestive that the mass
for the bound states of these four types of wrapped D-branes
takes the form
\eq{
  \label{masscontra}
      M_1^2={M^2_{\rm pl}\over t_1} \Bigg[&\bigg( {n_1
            +b_3 n_2 + b_2 n_3 + b_2 b_3 n_4\over  t_2^{1\over
              2} \,t_3^{1\over 2}  }\bigg)^2+
        \bigg( { (n_2 + b_2 n_4) \,   t_3^{1\over
              2} \over t_2^{1\over 2}  }\bigg)^2+\\[0.2cm]
        &\bigg( { (n_3 + b_3 n_4) \,   t_2^{1\over
              2} \over t_3^{1\over 2}  }\bigg)^2+
         \bigg( { n_4    t_2^{1\over
              2} \, t_3^{1\over 2}  }\bigg)^2\Bigg]\,.
        }
The $n_i$ denote the four wrapping numbers of the wrapped $D0$, $D2$
and  $D4$ branes.

Similarly, the other four wrapped $D2$-branes from table
\ref{table_Dbranes}  are also related via turning on the
(quasi-)axions $v_2$ and $v_3$ so that we propose
\eq{
  \label{masscontrb}
      M_2^2={M^2_{\rm pl}\over t_1} \Bigg[&\bigg( {p_1
            +v_3 p_2 + v_2 p_3 + v_2 v_3 p_4\over  u_2^{1\over
              2} \,u_3^{1\over 2}  }\bigg)^2+
        \bigg( { (p_2 + v_2 p_4) \,   u_3^{1\over
              2} \over u_2^{1\over 2}  }\bigg)^2+\\[0.2cm]
        &\bigg( { (p_3 + v_3 p_4) \,   u_2^{1\over
              2} \over u_3^{1\over 2}  }\bigg)^2+
         \bigg( { p_4    u_2^{1\over
              2} \, u_3^{1\over 2}  }\bigg)^2\Bigg]\,.
}
So far, the arguments behind the mass formulas \eqref{masscontra}
and \eqref{masscontrb} were still fairly standard.
Next, we are dealing with the two KK-modes
\eqref{largetKKs} along the two one-cycles of the first $T^2$ factor and the corresponding 
two wrapped NS5-brane states listed in table \ref{table_NSbranes}.
 For these final  four states we apply just the analogy to the
fundamental string.
For the fundamental string, a wrapped string along the $x(y)$-cycle
in the presence of a Kalb-Ramond  field $B_2$ changes the definition
of the canonical momentum and provides a correction to the
mass of the KK-mode along the $y(x)$-direction.
Analogously, we now propose that a wrapped NS5-brane
along the $x(y)$-cycle
in the presence of the magnetic dual $B_6$ field (with all legs
along $T^6$) changes the definition
of the canonical momentum and provides a correction to the
mass of the KK-mode along the $y(x)$-direction.
Applying this logic we obtain the mass formula
\eq{
  \label{masscontrc}
      M_3^2={M^2_{\rm pl}\over t_1} \Bigg[&\bigg( {m_1
            +v_1 m_2 + \rho\, m_3 + v_1 \rho\, m_4\over  \sigma \,u_1^{1\over 2}  }\bigg)^2+
        \bigg( { (m_2 + \rho\, m_4) \,   u_1^{1\over
              2} \over \sigma  }\bigg)^2+\\[0.2cm]
        &\bigg( { (m_3 + v_1 m_4) \,   \sigma  \over u_1^{1\over 2}  }\bigg)^2+
         \bigg( { m_4    u_1^{1\over
              2} \, \sigma  }\bigg)^2\Bigg]\,.
}
Recall that $\rho$ is the magnetic dual of the Kalb-Ramond field $B_2$
with both legs along the 4D space-time. Therefore, it is $B_6$ with
all legs along the 6 toroidal directions.
Finally, we put all these the contributions together and also add
the contribution from the oscillators of the low-tension 4D string
arising from the wrapped NS5-brane
\eq{
          M^2=M_1^2+M_2^2+M_3^2 + {M^2_{\rm pl}\over t_1} \kappa^2 N\,.
 }
 Thus, we claim that all two KK-modes and the many wrapped brane states can form
 bound states whose mass will be given by this relation.
 This is certainly a strong claim, but the analogy to the weak
 coupling  limit is striking.
 
 Analogous to the mass expression \eqref{masstower},  introducing now continuous variables, for the
 derivative with respect to the axion $\rho$ we obtain
\eq{
 \partial_{\rho} M \simeq {M_{\rm pl}\over \sqrt{t_1}} {1\over 2\, r\,
  \sigma^2}\Big( 2\, w_3\, y_3  +2\, x_3\, z_3\Big) \,.
}
Hence, compared to the results in the perturbative limit \eqref{allderiv}, we notice the
factor $\sigma^2$ (instead of $\sigma$). This will correctly
reproduce the  term $d\rho^2/(4\sigma^4)$ in the
tree-level metric \eqref{modulimetric}.


\clearpage

\bibliography{references}  
\bibliographystyle{utphys}


\end{document}